\documentclass{aa}

\pdfoutput=1
\usepackage[english]{babel}
\usepackage{natbib}
\usepackage{amsmath,amsfonts,amssymb,textcomp}
\usepackage{amssymb}
\usepackage{pdflscape}
\usepackage{multirow}
\usepackage{varioref}
\usepackage{hyperref}
\usepackage{psfig}
\usepackage{graphicx,graphics,subfigure}
\usepackage[usenames,dvipsnames]{xcolor}
\bibpunct{(}{)}{;}{a}{}{,}
\hypersetup{backref,
colorlinks=true,citecolor=blue, urlcolor=magenta}

\labelformat{equation}{\textup{(#1)}}

\labelformat{Fig.}{\textup{(#1)}}

\begin{document}
\graphicspath{{figures/}}

\title{Star formation and black hole accretion activity in rich local clusters of galaxies}
\titlerunning{SF and BH accretion activity in rich local clusters of galaxies }

\author{Matteo Bianconi\inst{1}, Francine R. Marleau\inst{1}, Dario Fadda\inst{2}}

\institute{Institut f\"{u}r Astro und 
Teilchenphysik, Leopold Franzens Universit\"{a}t Innsbruck,  Technikerstra\ss e 25/8, A-6020 Innsbruck, Austria \\ \email{matteo.bianconi@uibk.ac.at} \and
Instituto de Astrofisica de Canarias, E-38205 La Laguna, Tenerife, Spain;\\
Universidad de La Laguna, Dpto. de  Astrofísica, E-38206 La Laguna, Tenerife, Spain}

\authorrunning{M. Bianconi, F.R. Marleau, D. Fadda}

\abstract
{We present a study of the star formation and central black hole accretion activity of the
galaxies hosted in the two nearby (z$\sim$0.2) rich
galaxy clusters Abell 983 and 1731.} 
{We are able to quantify both the
obscured and unobscured star formation rates, as well as the presence of
active galactic nuclei (AGN) as a function of the environment in which the galaxy is located. } 
{We targeted the clusters with unprecedented deep infrared Spitzer observations
(0.2 mJy @ 24 micron), near-IR Palomar imaging  and optical WIYN spectroscopy. The
extent of our observations ($\sim$ 3 virial radii) covers the vast range of possible
environments, from the very dense cluster centre to the very rarefied
cluster outskirts and accretion regions.}
{The star forming members of the two clusters present star formation rates comparable with those measured in coeval field galaxies. The analysis of the spatial arrangement of the
spectroscopically confirmed members reveals an elongated distribution for A1731 with respect to the more uniform distribution of A983. The emerging picture is compatible with A983 being a fully
evolved cluster, in contrast with the still actively accreting A1731. }
{The analysis of the specific star formation rate reveals evidence of on-going galaxy pre-processing along A1731's filament-like structure. Furthermore, the decrease in the number of star forming galaxies and AGN towards the
cluster cores suggests that the cluster environment is accelerating the ageing process of the galaxies and blocking further accretion of the cold gas that fuels both star formation and black hole accretion activity.}
 

\maketitle
\section{Introduction}

 \begin{figure*}
 \centering
 \includegraphics[width=0.4\linewidth, keepaspectratio]{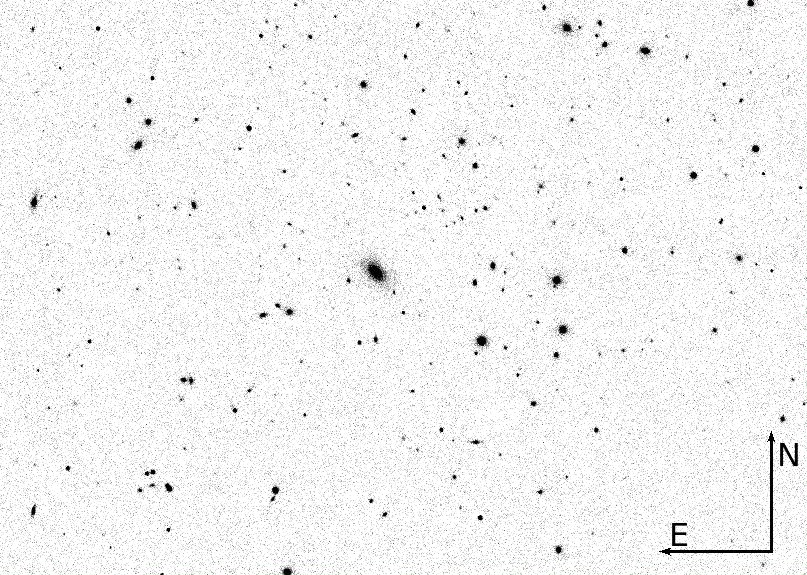}
 \includegraphics[width=0.4\linewidth, keepaspectratio]{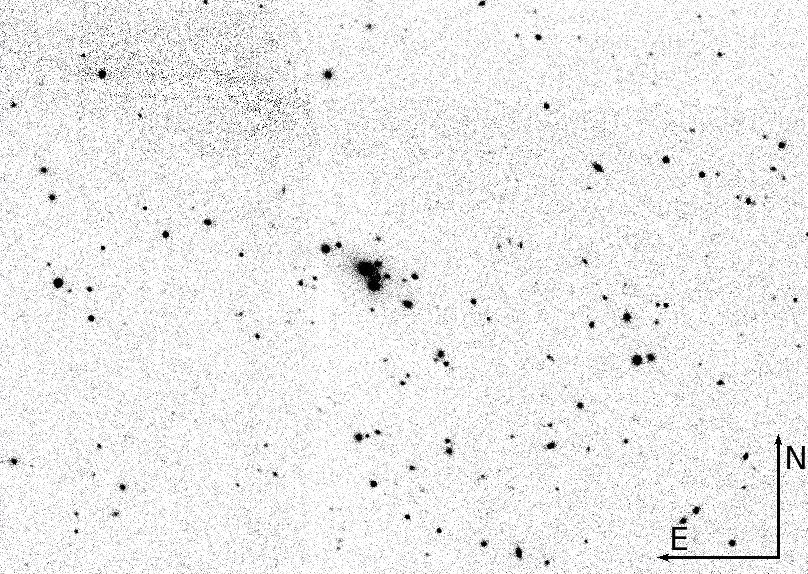}
\caption{WIRC Ks images of A983 (left panel) and A1731 (right panel), zoomed on the central $7' \times 5'$ region. The figures are centered on the brightest cluster galaxies.}
 \end{figure*}\label{bcg_wirc}
 
 \begin{table*}
\begin{center}
\begin{tabular}{lccccc}
\hline
Cluster & $\rm M_{200}$ [$\rm M_{\odot}$] & $\rm r_{200}$ [$\rm Mpc$] & $\sigma_{v}$ [$\rm km\, s^{-1}$] & $\bar{z}$ & Members\\
\hline
\hline
Abell 983 & 1.36 $\times 10^{15}$& 2.140 & 1071 &  0.20 & 134\\
Abell 1731 & 1.92 $\times 10^{15}$ &2.408 &  1201 &  0.19 & 91\\
\hline
\end{tabular}
\end{center}
\caption{Main properties of the observed clusters.}\label{cluster_props}
\end{table*}
The current paradigm of structure formation predicts that the galaxy population in clusters is evolving as new members are accreted from the surrounding field region \citep{balogh98, vogelsberger14}. The
dependence of the galaxy evolution on the environment in which they are located is proven to
be tight. The morphology-density relation \citep{dressler80} implies that
the environment affects the star formation history, color, and
structure of galaxies. As a result, the young and active galaxies can
be found typically in the cluster outskirts, and the passive ones in the
cluster core.  The processes concurring to this fast ageing of the
galaxies are several.  The ram pressure due to the intracluster
medium (ICM) \citep{gunn72, steinhauser12} builds on the gas present in the
galaxy, as it travels through the cluster. The consequent compression
of the galaxy gas leads to sudden enhancement of the star
formation (SF) but it also favours the gas removal at later stages. Galaxies can also suffer gas losses via gravitational disturbance.  \citet{larson80} noticed that the hot gaseous halo of the galaxy is stripped as it enters the cluster environment. This process, called galaxy strangulation,  prohibits further accretion of gas on the galaxy. Furthermore,  the high density of galaxies in the
cluster environment promotes frequent gravitational
encounters, that induce the so-called harassment process. The
dynamical equilibrium of the gas is altered and its collapse is
facilitated, due to the perturbation of the gravitational potential. This leads to new bursts of SF and hence further ejection of portion of the remaining gas, due to stellar winds.
The efficiency of gravitational encounters to trigger new SF episodes
increases in dense environments with low velocity dispersion. Such conditions are found in
the filaments, along which the galaxies are funneled and accreted in
the cluster (\citealt{balogh00}; \citealt{diaferio01};
\citealt{okamoto03}).  For example, \citet{fadda08} and \citet{biviano11} found a
higher fraction of star forming galaxies in the filament of Abell 1763, double with
respect to the cluster core and outskirts.
\citet{wolf09}, \citet{biviano11}, \citet{delucia12},
\citet{wetzel13} showed that long lasting encounters are more likely to trigger SF in filaments than abrupt processes such as mergers(see also
\citealt{wijesinghe12}). \citet{haines15} found the specific star
formation rate (sSFR) of massive galaxies in a sample
of 30 cluster to be $\rm \sim 30\%$ lower than their counterparts
in the field. A subsequent modelling allowed to constrain the
quenching time scale in the range of $\rm 0.7-2.0$ Gyr, consistent
with the characteristic accretion time scale of galaxies in
clusters. \cite{peng15} support the scenario in which
local quiescent galaxies with stellar masses smaller than $\rm10^{11}
\,M_{\odot}$ (i.e. the vast majority of galaxies) are primarily
quenched as a consequence of strangulation.  To correctly interpret the evolution of galaxies and their accretion history, it is therefore
fundamental to cover observationally the entire extent of galaxy
clusters, with the inclusion of the outskirts and the possible accretion structures. In addition, it is important to consider clusters that
do not present evidences of ongoing mergers. The intense disturbance
on the cluster dynamics can hide or cancel the effects of the
environment on galaxies and prohibit the study of the secular
accretion of the galaxies.  A fundamental
difficulty in studying these objects is the contamination of
interlopers, i.e. foreground and background galaxies. This issue
becomes more important at larger clustercentric radii where the
relative fraction of interlopers is larger. Therefore,
spectroscopic redshift measurements are required.

\begin{figure*}
 \centering
  \includegraphics[width=0.38\linewidth, keepaspectratio]{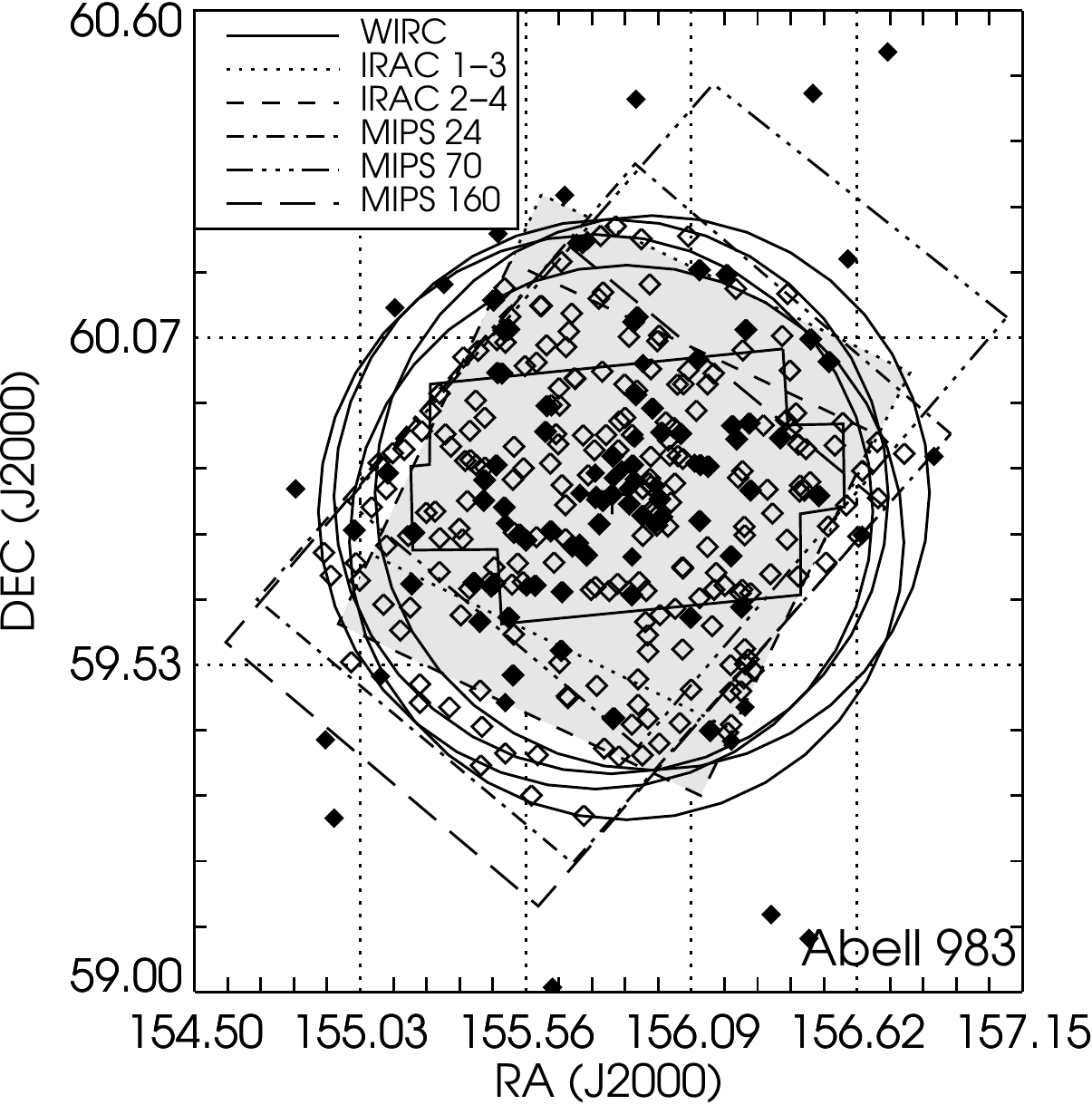}\includegraphics[width=0.51\linewidth]{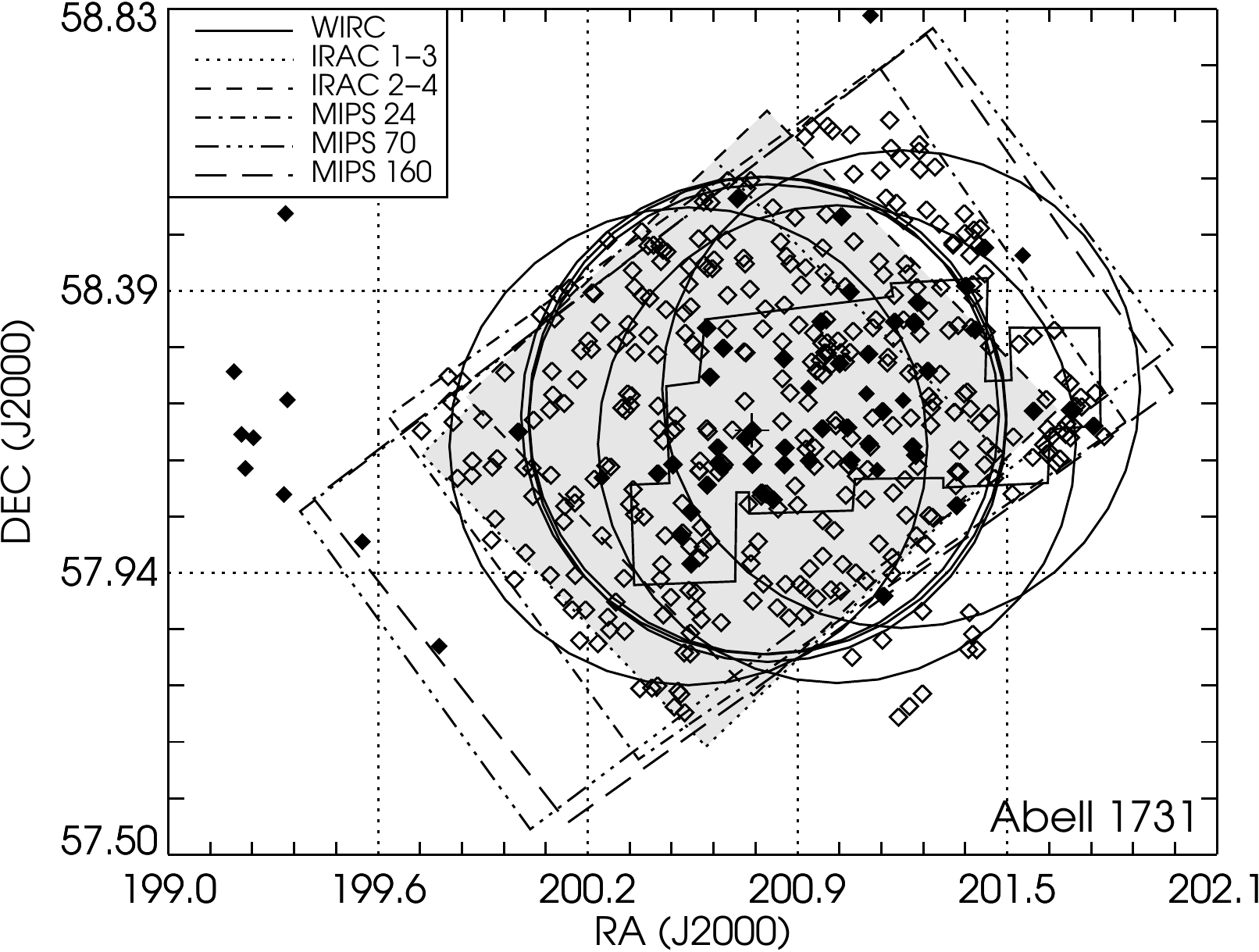}
\caption{Coverage of IR and optical data for A983 (left panel)
  and A1731 (right panel). The open and filled diamonds
  correspond to the HYDRA observed targets and the spectroscopically
  confirmed members, respectively. The solid line polygon delineates
  the WIRC observations. The shaded region marks the IRAC fields. The
  3.6 and 5.8 $\rm \mu$m data are outlined with a dotted line, and the
  4.5 and 8.0 $\rm \mu$m data are outlined with a dashed line.  The
  footprints of the MIPS 24 $\rm \mu$m, 70 $\rm \mu$m and 160$\rm
  \mu$m field are delimited by the dot-dashed, dash-triple dot, and
  long dashed line, respectively. The large circles mark the WIYN/Hydra field of view.}\label{coverage}
 \end{figure*}

The star formation rate (SFR) is an instantaneous quantity, directly susceptible to the influence of external processes, and therefore well suited for the study of environmental effects (\citealt{fadda02}). Robust
measurements of the SFR are necessary, and hence must include both the obscured (via the infrared emission) and unobscured (via the UV and optical emission) star formation activity. In addition, recent studies correlate the presence and the characteristics of
active galactic nuclei (AGN) to their parent galaxies as
well as to the environment in which these galaxies are
located. Contradictory scenarios have been proposed so far in the
literature, without any clear predominance \citep{sabater13}. The
cluster environment allows us to study the duality of the
environmental effects: on the large scale, influencing the gas supply, and on the local scale, regulating the accretion of the AGN via galaxy-galaxy interactions.

The galaxy clusters Abell 983 and 1731 (hereafter A983 and A1731, respectively) are local rich galaxy clusters that we selected as targets for our study.
ROSAT X-ray images of A983 reveal the uniform emission of
the ICM, that traces the relaxed and virialised state of the
cluster. On the other hand, A1731 shows signs of a less homogeneous X-ray surface brightness. However, the shallow depth of the X-ray data does not allow to draw secure conclusions on distribution of ICM. With respect to A983, A1731 presents a higher density of galaxies in the core, with two bright cluster galaxies which are surrounded by smaller objects (see Figure~\ref{bcg_wirc}).

In this paper, we present the study of star formation and black hole accretion activity in A983 and A1731. This study is based on deep infrared {\em Spitzer} observations, near-IR Palomar imaging and optical WIYN spectroscopy. This paper is structured the following way. In Section~\ref{data_set}, we
present in detail the observational program and the data reduction. In Section~\ref{cat}, we present our matched photometric and spectroscopic catalogue.
 In Section~\ref{data_analysis}, we present the results of the analysis of our
dataset.  In Section~\ref{results}, we discuss the scenario that can be drawn 
from our results. In Section ~\ref{summary}, we summarise the results and 
present future prospects of the project. Throughout this paper, 
we assume $\rm H_0= 70\, km \,s^{-1} Mpc^{-1}$, $\rm\Omega_M= 0.3$ 
and $\rm\Omega_{\Lambda}= 0.7$. At the
clusters' redshift, 1 arcsec corresponds to $\rm \sim 3.2\,kpc$.

\section{The data set}\label{data_set}
The main quantitative properties of
A983 and A1731 are obtained in Section~\ref{members} and summarized in
Table~\ref{cluster_props}.
The area covered by our photometric and spectroscopic observations is
shown in Figure~\ref{coverage}. The main details of these observations are summarised in Table~\ref{obs_a983} and Table~\ref{hydra_obs}. In the following sections, we introduce and describe the
observing strategy and the subsequent data reduction and analysis that
led to the production of the cluster members catalogues.

\subsection{Observations: mid to far-IR with Spitzer IRAC and MIPS}
A983 and A1731 were observed as part of the Spitzer program 20512 (PI: Dario Fadda). A983 and A1731 were selected
along with another cluster, Abell~1763 (\citealt{edwards10}), as being
rich systems and located at a similar redshift  in
regions with extremely low Galactic emission, allowing us to measure the SF from the infrared
with a lower limit of $\rm\sim 1\;M_{\odot}\,yr^{-1}$ .  Furthermore, these clusters were targeted for being at low redshift ($\rm z\sim0.2$), allowing a wide coverage that extends to approximately to 3 virial
radii.

The Spitzer IRAC images were taken with the instrument set in mapping
mode, allowing the simultaneous imaging of the field of view in the 4
different channels corresponding to 3.8, 4.5, 5.8 and 8 $\mu$m. Each
pointing of the telescope was dithered three times to allow for the
detection and removal of transient phenomena, such as cosmic rays. The
IRAC images cover 39.2$\times$39.2 sq. arcmin. on the plane
of the sky, corresponding to 7.3$\,$Mpc$\times$7.3$\,$ sq. Mpc. The MIPS images were taken using slow telescope scanning, suitable for
covering large portions of the sky. Throughout the scanning of the
sky, the subsequent frames overlap each other, for a more efficient removal
of cosmic rays.  The motions of a secondary mirror in a cryogenic
bath, compensate for the oscillations of the telescope during the
scanning. The MIPS images overlap the region covered by the IRAC
images, and extend beyond that, reaching a size of 40.3
$\times$55.6 sq. arcmin. (7.5$\times$10.4 sq. Mpc). In the case
of A1731, the MIPS 70 and 160 $\rm \mu$m present a larger final
footprint ( $\rm \sim 2880'^2$ and $\rm \sim 2700'^2$,
respectively), due to a shift of $\sim$18' that
occurred to the pointing of the last Astronomical Observational
Request (AOR). In the 24 $\mu$m band, we achieved a flux limit of 0.2
mJy in 29.9 ks.  Table~\ref{obs_a983} summarises the main properties of our near-
to far-IR dataset.  The Full Width at Half Maximum (FWHM) of the Point Spread Function (PSF) for the
Spitzer observations is quoted from the IRAC and MIPS Data
Handbooks\footnote{\label{note1}http://irsa.ipac.caltech.edu/data/SPITZER/docs/sitemap/}.

\begin{figure*}
 \begin{center}
 \hspace{-0.2cm}\includegraphics[width=0.3\linewidth, keepaspectratio]{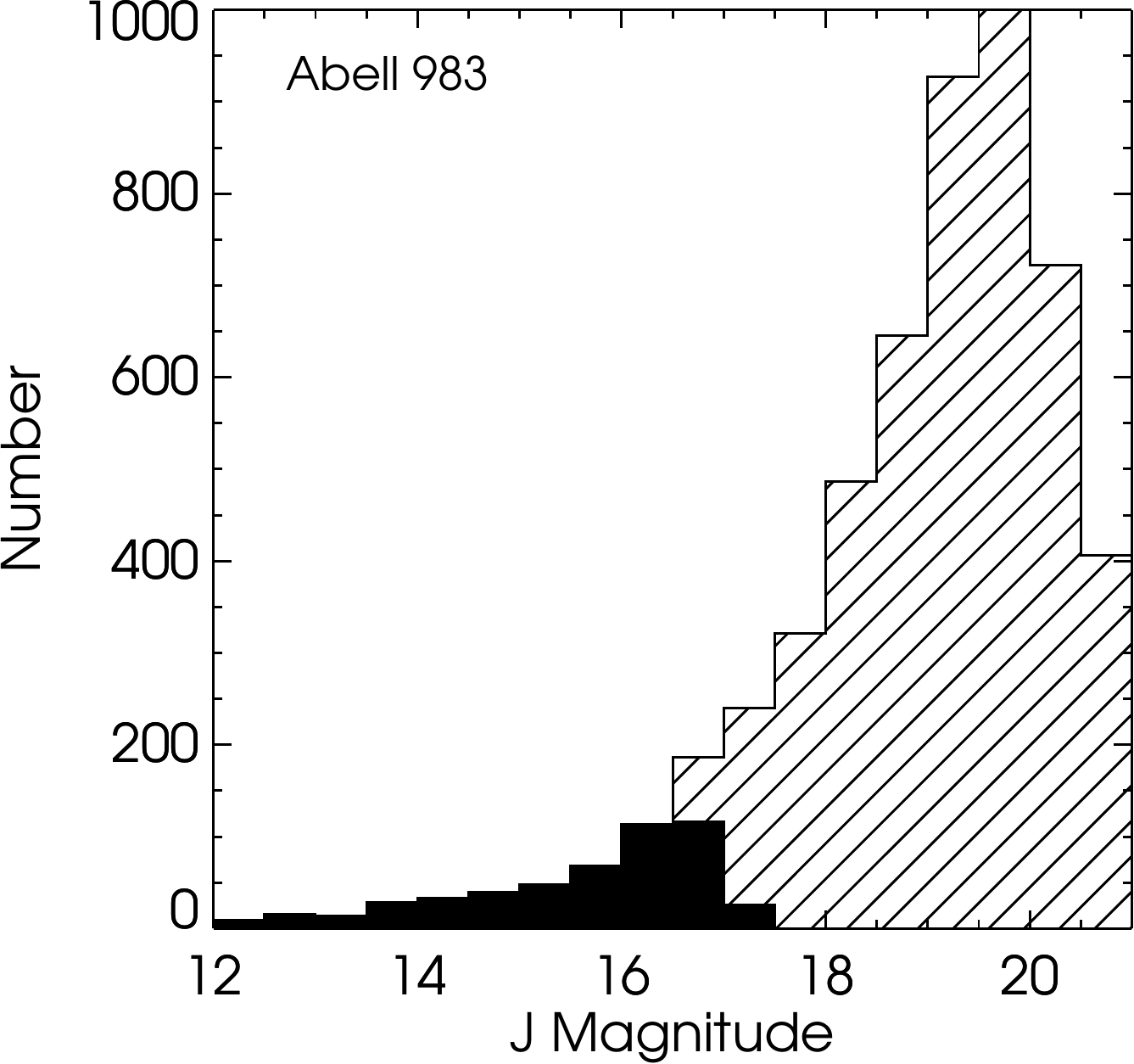}\includegraphics[width=0.295\linewidth, keepaspectratio]{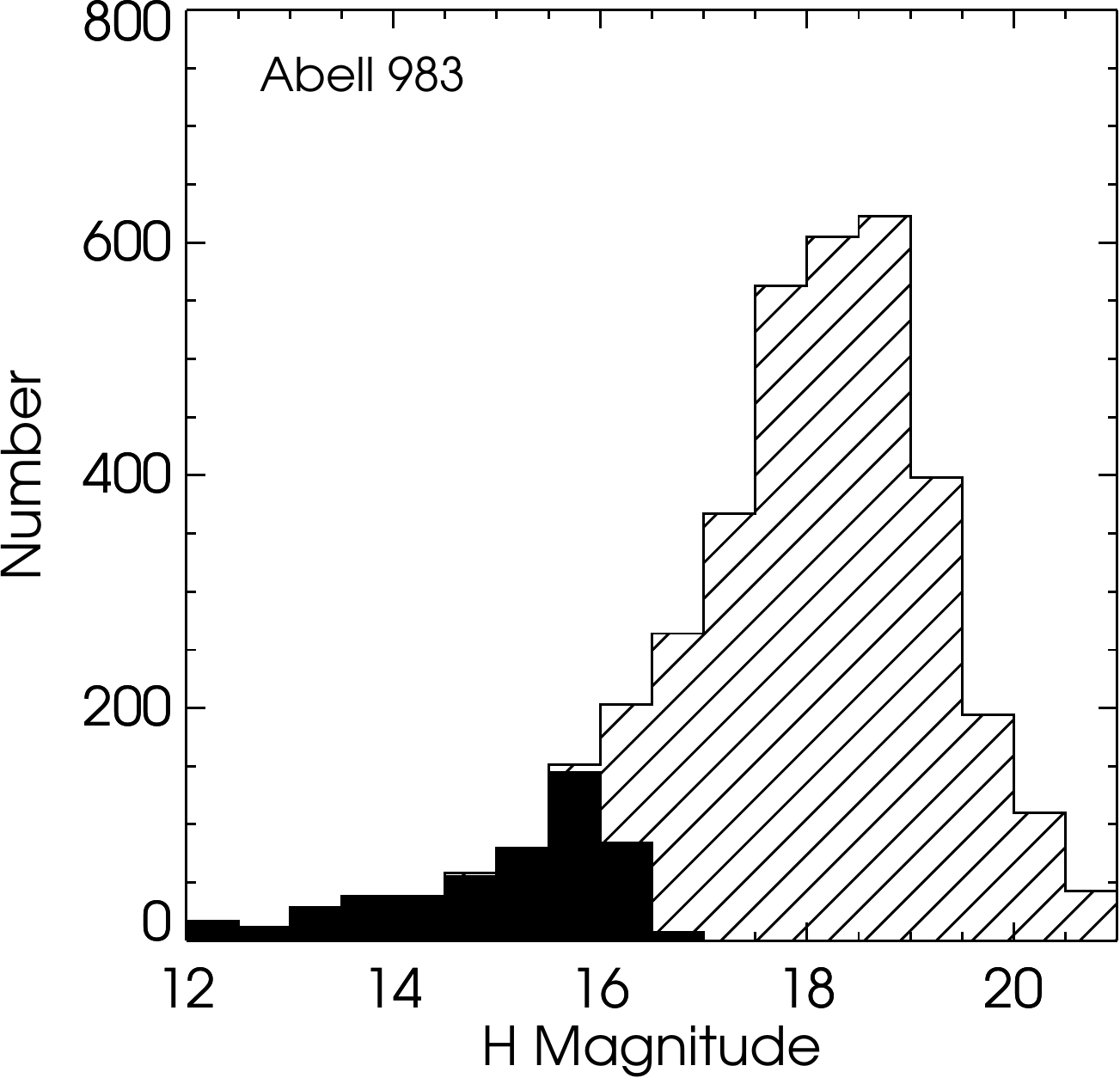}\includegraphics[width=0.3\linewidth, keepaspectratio]{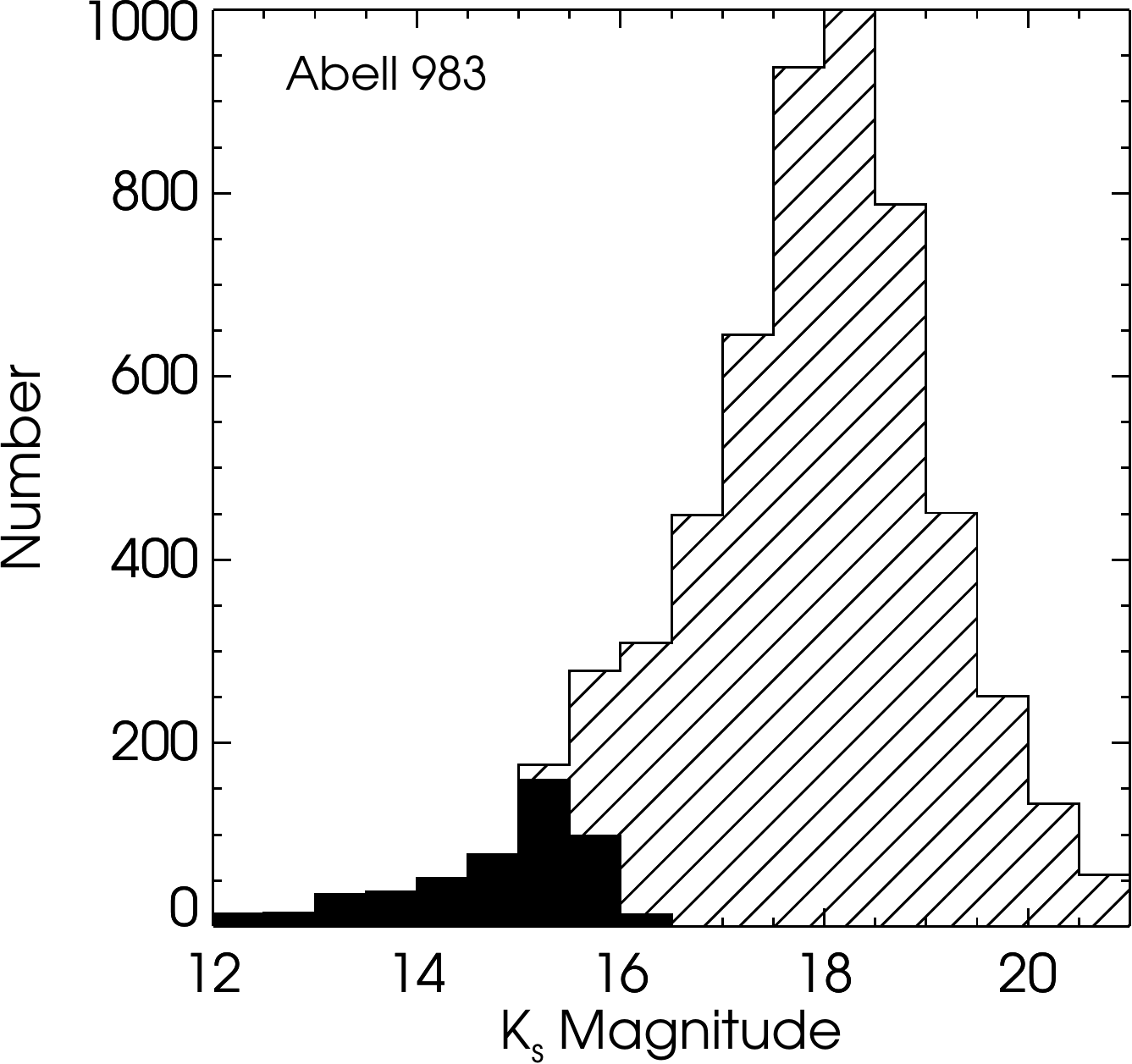}
 \includegraphics[width=0.3\linewidth, keepaspectratio]{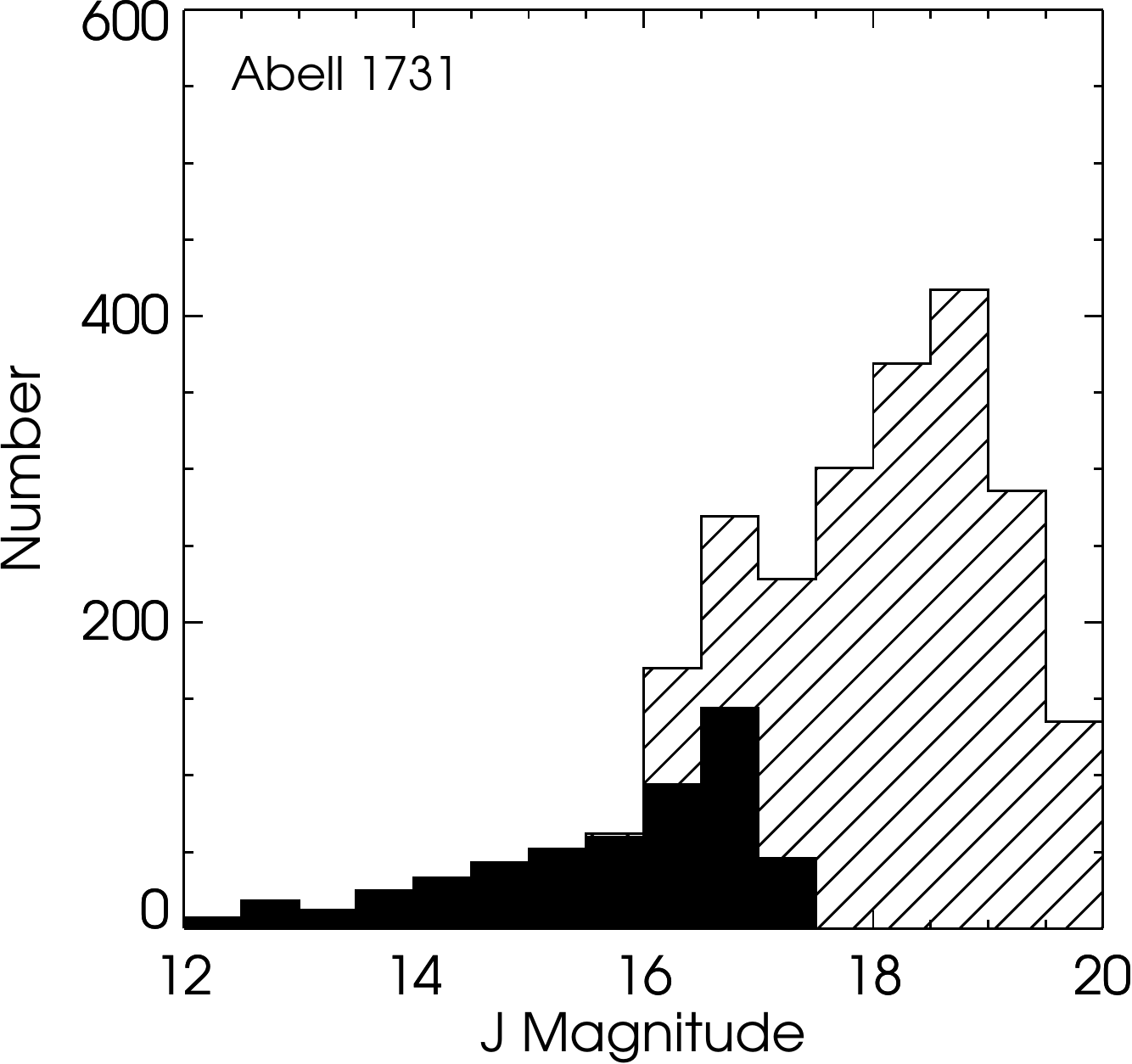}\includegraphics[width=0.3\linewidth, keepaspectratio]{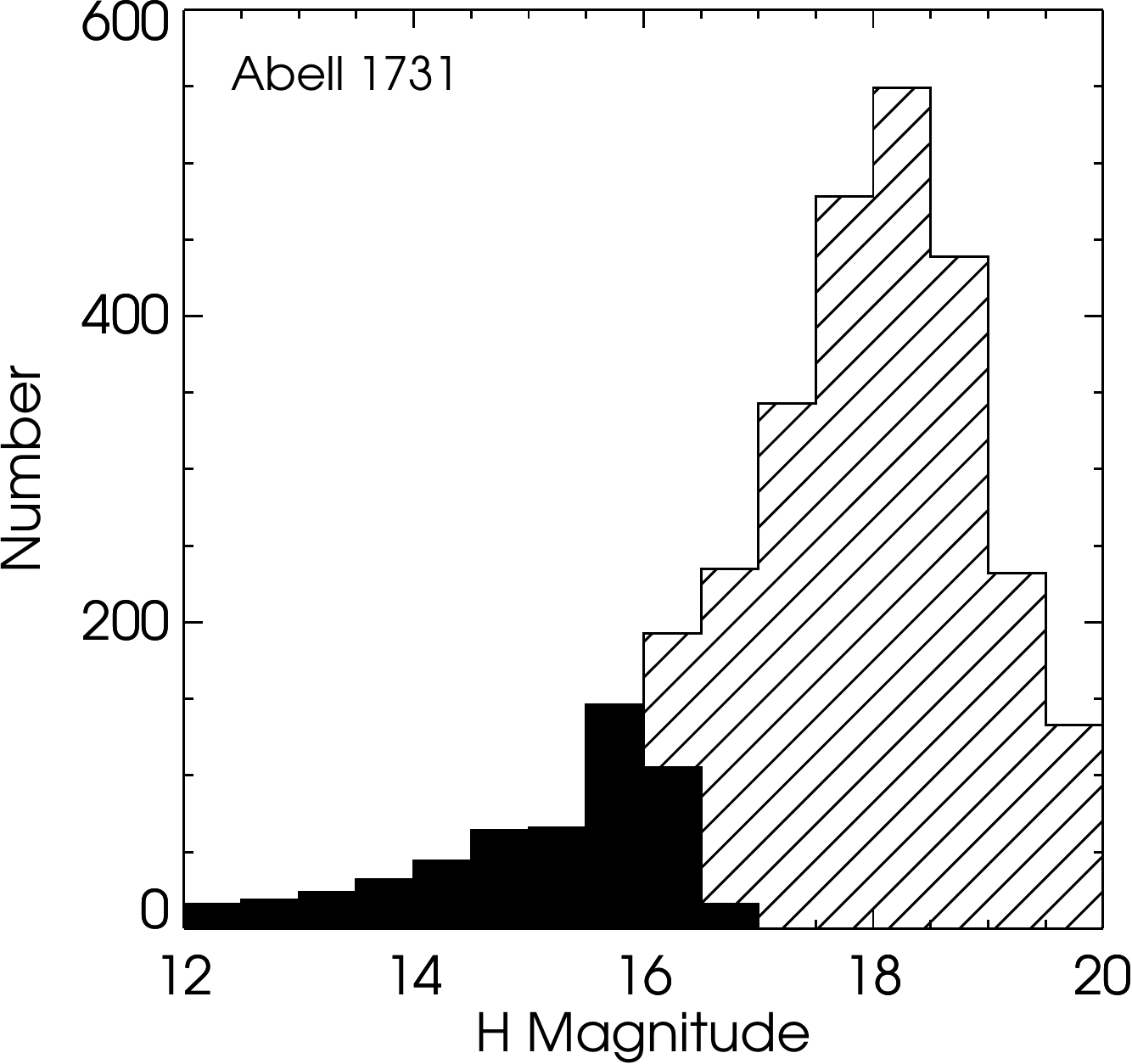}\includegraphics[width=0.3\linewidth, keepaspectratio]{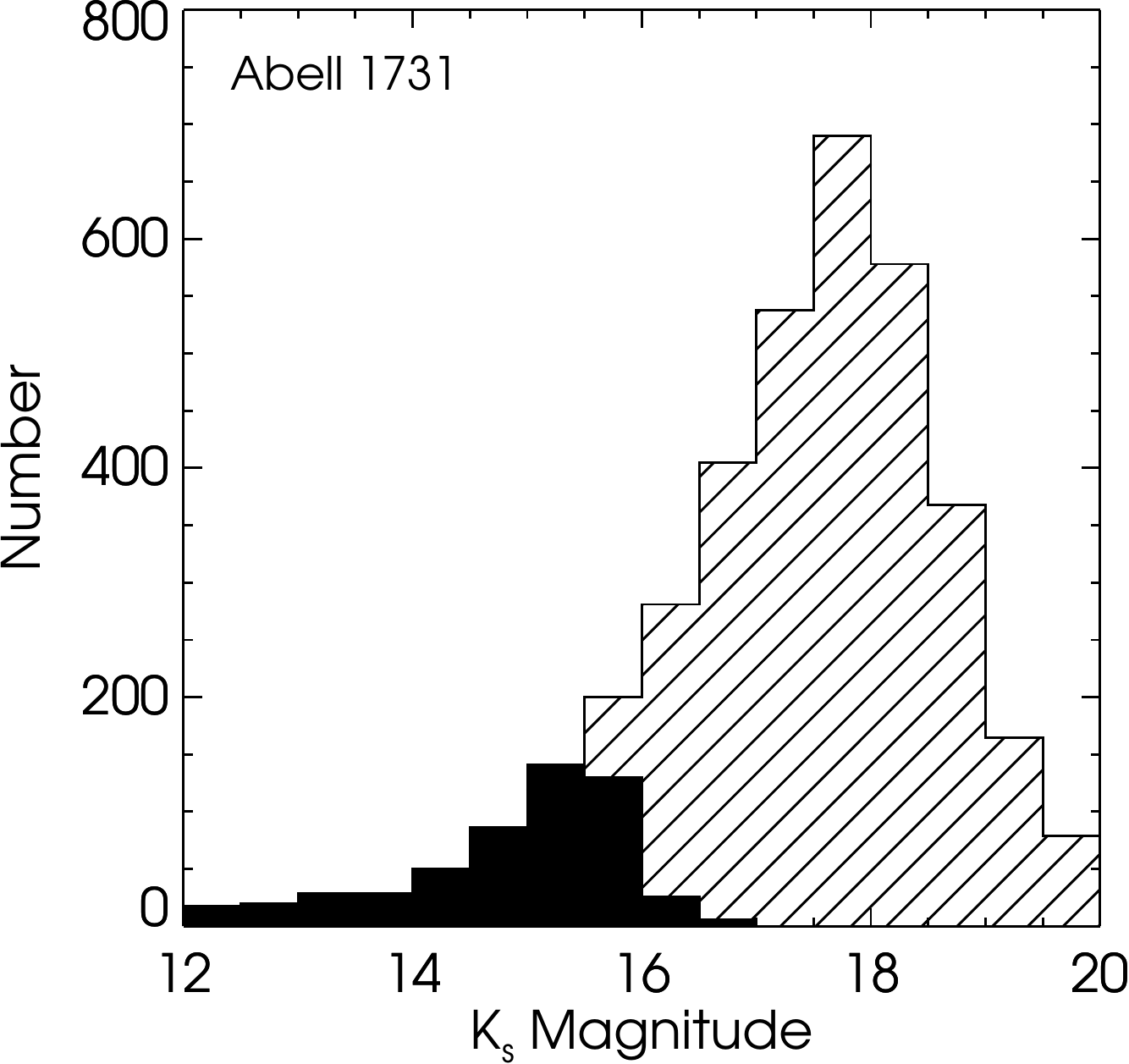}
\caption{The WIRC J, H, $\rm K_{s}$ number counts for A983 (top) and A1731
  (bottom). The histograms show the number of sources per bin of
  magnitudes J, H, and Ks from left to right, respectively. The
  overplotted filled histograms show the depth of the archival 2MASS
  data in the same region of the sky covered by the WIRC observations.}\label{palomar_histo}
   \end{center}
 \end{figure*} 
The standard Spitzer pipeline processes the raw data, outputting basic
calibrated datasets (BCDs). The corrections that were applied
include dark subtraction, cosmic ray correction, detector
linearization, flat field application and muxbleeding correction\footnote{http://irsa.ipac.caltech.edu/data/SPITZER/docs/\\irac/iracinstrumenthandbook/1}. The latter artifact consists
  of an electronic ghosting that can appear on the detector due to the
  delay of the detector in returning to its ground state, e.g. after
  the read out of a bright source. We applied additional corrections
to the BCDs, in order to obtain a higher signal-to-noise (SNR) in the images
and to avoid false source extraction when running automated software
for object detection. The additional corrections follow the procedure
described in \citet{fadda06} and \cite{edwards10}. For the
  IRAC bands, we corrected for column pulldown, jailbars, stray light
  and spurious effects of bright sources. We then applied a superflat to the IRAC channel 3 and 4 BCDs. Specifically, each BCD, after masking the bright sources,  was divided by its median value. Then, the median value of all these  BCDs was computed and each original BCD was divided
 by this superflat. This procedures helps in removing background gradient present in the final mosaic. For the MIPS data, we corrected
  for jailbars, background discontinuities, and pixel distortions. We
  also applied a superflat and an additional flat accounting for the impurities on the  cryogenic scan mirror. The MIPS astrometry was corrected by extracting the position of the sources in sets of 25 consecutive frames and matching them to the SDSS DR10 r'-band catalogue  \citep{ahn14}.

\begin{table*}
\begin{center}
\begin{tabular}{lcccccc}
\hline
Cluster& Instrument & $\lambda_{cent}$ [$\mu$m] &  Date & Time [min]& Coverage[$^{\prime 2}$] & FWHM of PSF [$^{\prime\prime}$]\\
\hline
\hline
Abell 983 & IRAC & 3.6 4.5 5.8 8.0 &   2005 Nov 26 & 72.9 (each) & 1600&	1.66 1.72 1.88 1.98\\
& MIPS & 24  &   2006 May 07 & 498.6 & 2200&	5.9\\
&MIPS & 70  &  2006 May 07 & 498.6& 2035&	16\\
& MIPS & 160 &  2006 May 07 & 498.6& 2000	&40\\
& WIRC J & 1.250 &  2007 Mar 26-27 & 61.0& 1000& 1.3\\
& WIRC H & 1.635 &  2007 Mar 26-27 & 99.3& 1000& 1.2\\
& WIRC Ks & 2.150 &  2007 Mar 26-27 & 67.2& 1000& 1.4\\

Abell 1731 & IRAC & 3.6 4.5 5.8 8.0 &   2005 Jun 13 & 72.7 (each) & 1600&	1.66 1.72 1.88 1.98\\
& MIPS & 24  &  05 Dec 08/06 Jun 12 & 498.6 & 2200& 	5.9\\
& MIPS & 70  &  05 Dec 08/06 Jun 12 & 498.6 & 2880& 16\\
& MIPS & 160 &  05 Dec 08/06 Jun 12 & 498.6 & 2700&	40\\
& WIRC J  & 1.250 &  2008 Apr 24-26 & 31.5& 950& 1.3\\
& WIRC H & 1.635  &  2008 Apr 24-26 & 31.5& 950& 1.2\\
& WIRC K$_{s}$  & 2.150 &  2008 Apr 24-26 & 41.0& 950& 1.4\\
\hline
\end{tabular}
\end{center}
\caption{Near to far infrared observations of A983 and A1731.}\label{obs_a983}
\end{table*}

\begin{table*}
\begin{center}
\begin{tabular}{lcccc}
\hline
Cluster & Pointing & Date & FOV Centre [R.A., Dec. (J2000)] & Integration time [sec] \\
\hline
\hline
Abell 983  & 1 & 2006 Apr 28 & 10:23:08.087 +59:46:58.00 & 3$\times$1200\\
 		   & 2 & 2006 Apr 29 & 10:23:31.913 +59:43:58.00 & 3$\times$1200\\
           & 3 & 2006 Apr 30 & 10:23:20.009 +59:48:28.00 & 3$\times$1200\\
           & 4 & 2008 Jan 14 & 10:23:51.566 +59:48:49.00 & 2$\times$2200\\
Abell 1731 & 1 & 2006 Apr 29 & 13:22:09.007 +58:08:32.16 & 3$\times$1200\\
           & 2 & 2006 Apr 28 & 13:23:05.127 +58:11:26.16 & 3$\times$1200\\
           & 3 & 2006 Apr 28 & 13:22:59.819 +58:10:44.16 & 3$\times$1200\\
           & 4 & 2006 Apr 29 & 13:23:05.129 +58:10:44.16 & 3$\times$1200\\
           & 5 & 2006 Apr 30 & 13:24:39.957 +58:13:56.16 & 3$\times$1200\\
           & 6 & 2006 Apr 30 & 13:23:54.371 +58:08:44.16 & 3$\times$1200\\
\hline
\end{tabular}
\end{center}
\caption{Details of the spectroscopic observations with HYDRA at WIYN.}\label{hydra_obs}
\end{table*}
The IRAC sources were extracted using SExtractor \citep{bertin96}. We measured
aperture fluxes using a radius of 3~arcsec for each source with a
SNR larger than 3.5 and multiplied the flux of each
channel by the corresponding point source aperture correction (see
Table~\ref{ap_corrections}), following \citet{surace05}. We computed aperture magnitudes using 9~arcsec radius for extended sources and we applied the extended sources aperture correction, following the instruction on the Spitzer's IRAC Handbook, for the galaxies  presenting an evident
extended
structure\footnote{\tiny http://irsa.ipac.caltech.edu/data/SPITZER/docs/\\irac/iracinstrumenthandbook/30/} ($\sim$1\% of the total number of sources).

The MIPS point sources were extracted using Starfinder
\citep{diolaiti00}. The software allowed us to estimate the PSF
directly from the image, accounting for the instrument design, using
an iterative procedure. The flux was then measured using apertures
with radii of 10, 16 and 20~arcsec for 24, 70 and 160
$\rm \mu m$, respectively. Multiplicative aperture corrections and color
corrections were applied accordingly (see
Table~\ref{ap_corrections})\footnote{http://irsa.ipac.caltech.edu/data/SPITZER/docs/\\MIPS/mipsinstrumenthandbook/50/}.
We checked for the presence of extended sources in the MIPS $24 \mu$m
by PSF fitting and removing the point sources from the original
image. We measured Petrosian fluxes of the extended sources ($\sim$2\% of the total number of sources) from the
residual image using SExtractor.

\subsection{Observations: near-IR with Palomar/WIRC}

Measuring the near-IR emission is essential in order to estimate the
stellar mass in galaxies. This is because most of the galaxy stellar
mass is locked up in the evolved population which emits most of its
light in the K-band \citep{kauffmann98}. Therefore, deep observations
in the near-IR are essential to obtain a robust estimate of the galaxy
stellar mass.  We obtained near-IR images of the central $\rm 60'\times 90'$
region of Abell 983 and Abell 1731 using the Wide InfraRed Camera (WIRC)
on the Palomar 200 inch telescope.

  A983 was imaged during two
  nights of observations on 2007 March 26-27. The images of A1731 were
  obtained during a second run on 2008 April 24-26. The nights were
  photometric, with seeing between 0.9 and 1.4" (see Table\ref{obs_a983}. Figure~\ref{bcg_wirc} shows the Ks images of the cluster central regions and  Figure~\ref{coverage} shows the footprints of the instrument.  The observing
  strategy for the two runs was different. To observe A983, we scanned
  the field by moving the field of view along in three strips with
  fixed declination. The J images were obtained with an exposure time
  of 40 s. In H and Ks, two coadds of 30 s and four co-adds of 10 s
  each, respectively, were taken to avoid saturation.  The total
  integration times for J, H, and Ks were 61, 99, and 67 minutes,
  respectively. For the second run, A1731 was imaged using a
  7-position dithering pattern centered on 9 different subfields.  In this case we used an exposure time of 30 s for J and H and
  three coadds of 13 s in the case of Ks. Hence, the total
  integration time was 31.5 minutes for J and H and 41 minutes for Ks.
  Dark frames were also obtained for each integration time.

The same data reduction technique was applied to all three bands using
a pipeline developed and kindly provided by Tom Jarrett.  A median dark
frame was subtracted from the data frames, and correction terms for
the flux nonlinearity were applied to correct the bias. A median sky made of a maximum of
10 frames was calculated and subtracted from the data frame before
flat fielding. The astrometry was checked using a list of known stars
  from the Two Micron All Sky Survey (2MASS;
  \citealt{skrutskie06}) and corrected by accounting for
  the rotational offset of the telescope  and compensating for the distortion of the
  WIRC instrument. SWarp \citep{bertin02} was used
to mosaic the frames together. Flux calibration was performed relative
to the 2MASS catalog. The magnitude limit that we aimed to reach for
these observations was $\rm M^*+2$, where $\rm M^*$ is the magnitude
at the knee of the luminosity function \citep{edwards10}. For a
cluster of galaxies at z$\sim0.2$, a typical value is
$\rm Ks^*= 15.6$ \citep{depropris03}.  The majority of the cluster
galaxies, being passive, present typical near-IR color indices of $\rm
J -Ks =1.7$ and $\rm H-Ks = 0.8$ \citep{fukugita95}. Hence, our near-IR
photometry reaches (at 3$\rm\sigma$) a depth of J$\sim19.5$, H$\sim18.5$,
and Ks$\sim17.5$, extending approximately two magnitudes deeper than
the 2MASS archival data (Figure~\ref{palomar_histo}).
 \begin{figure*}
 \centering
  \includegraphics[width=0.3\linewidth, keepaspectratio]{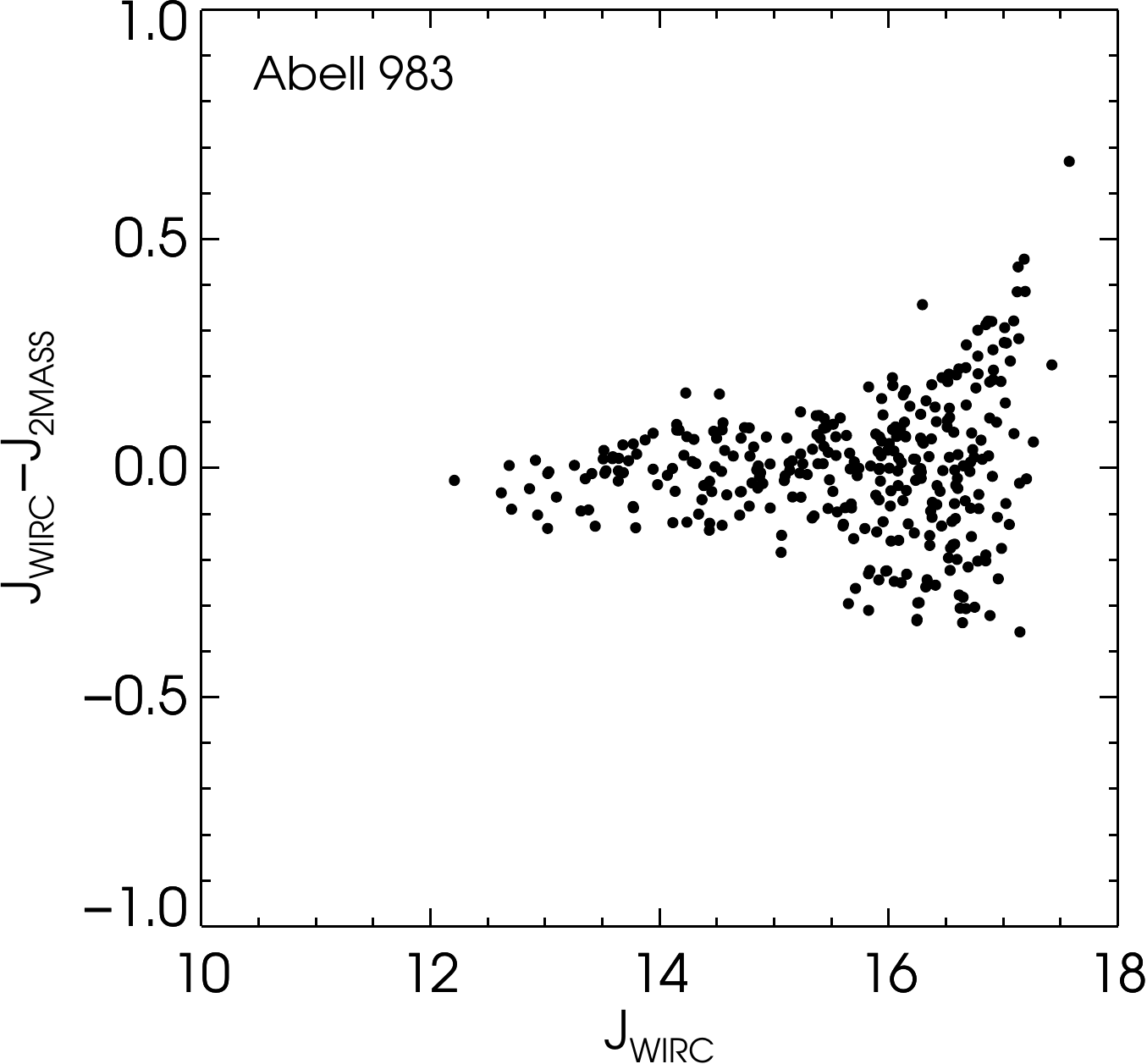}\includegraphics[width=0.3\linewidth, keepaspectratio]{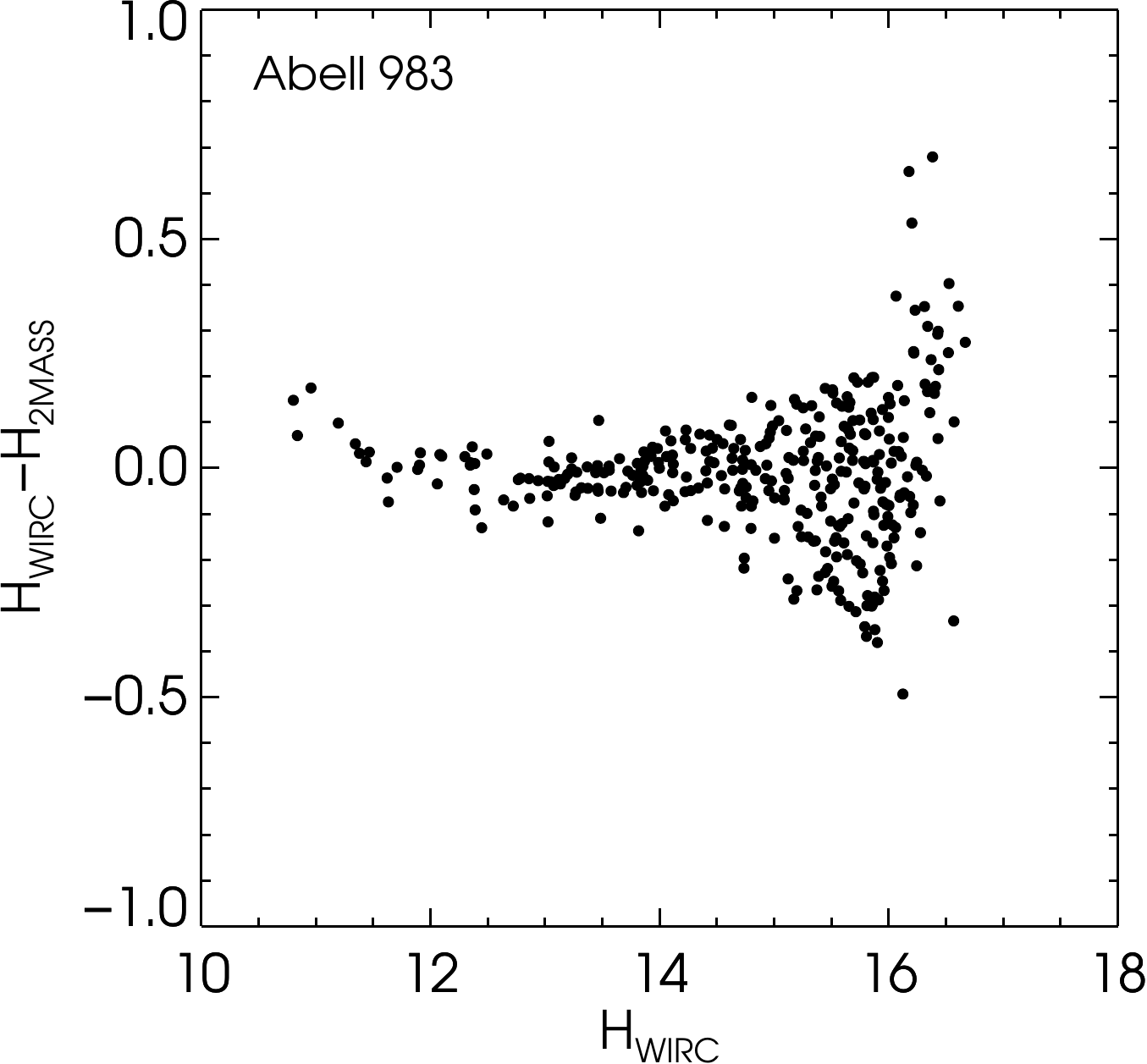}\includegraphics[width=0.3\linewidth, keepaspectratio]{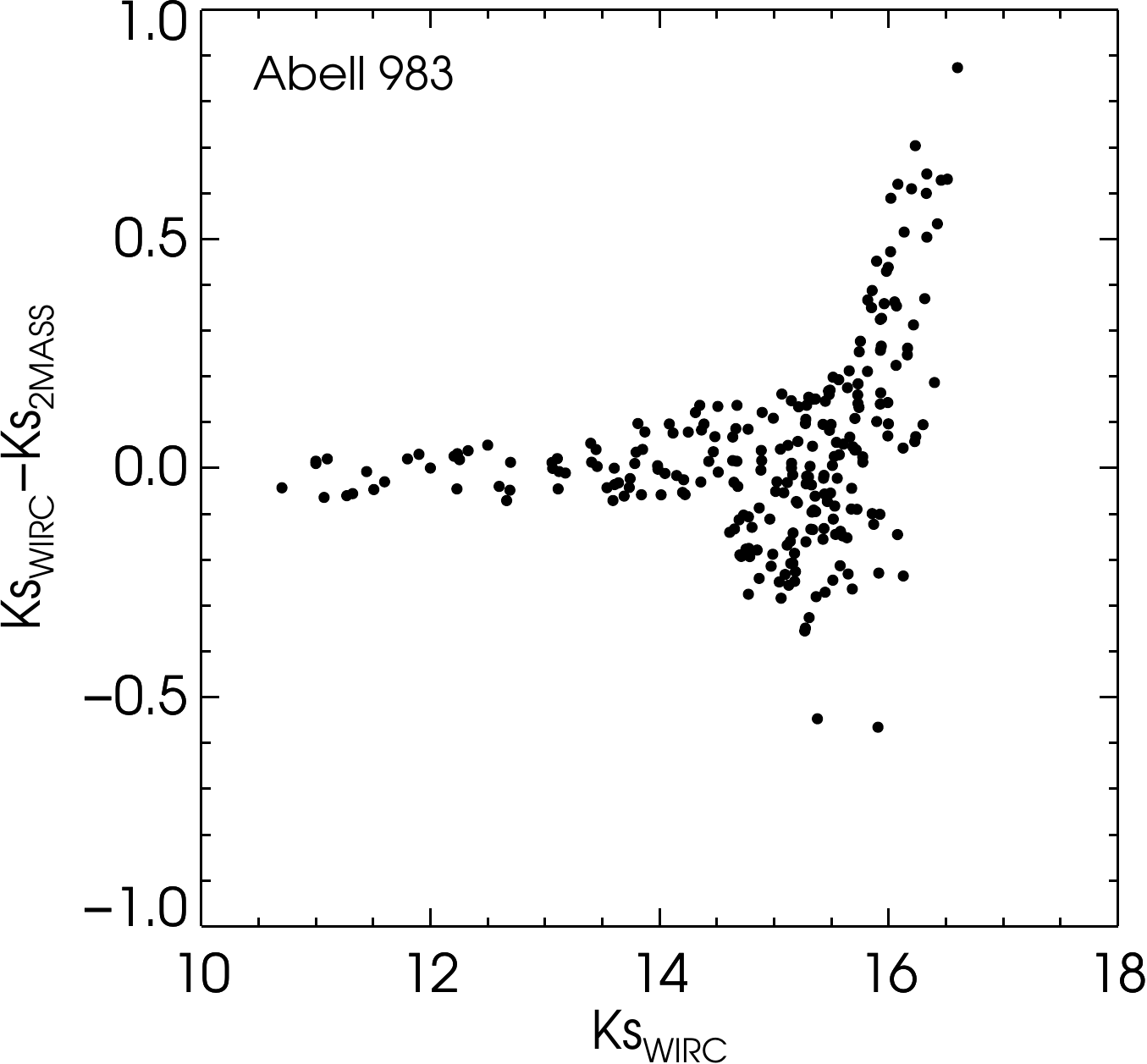}  
  \includegraphics[width=0.3\linewidth, keepaspectratio]{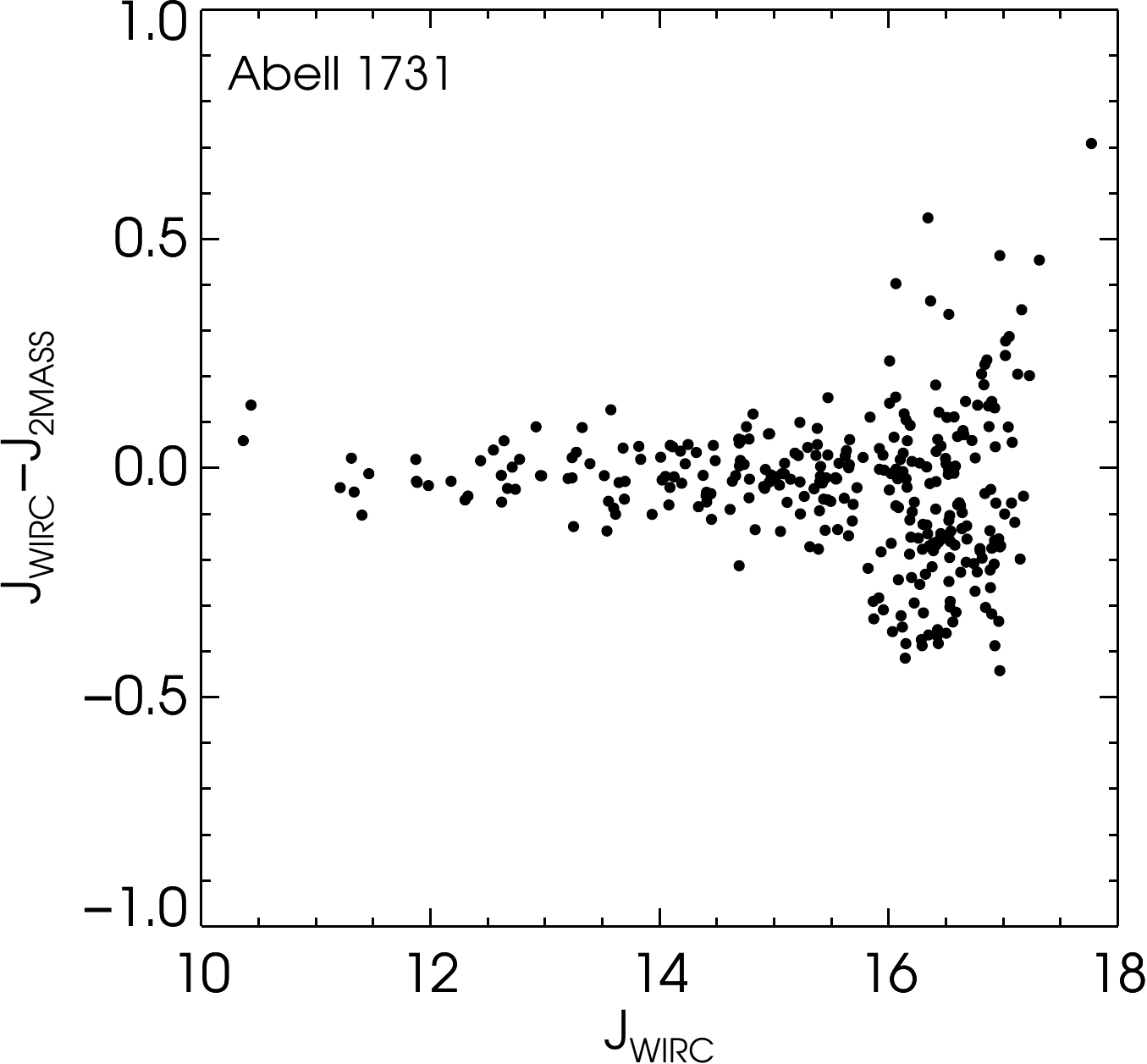}\includegraphics[width=0.3\linewidth, keepaspectratio]{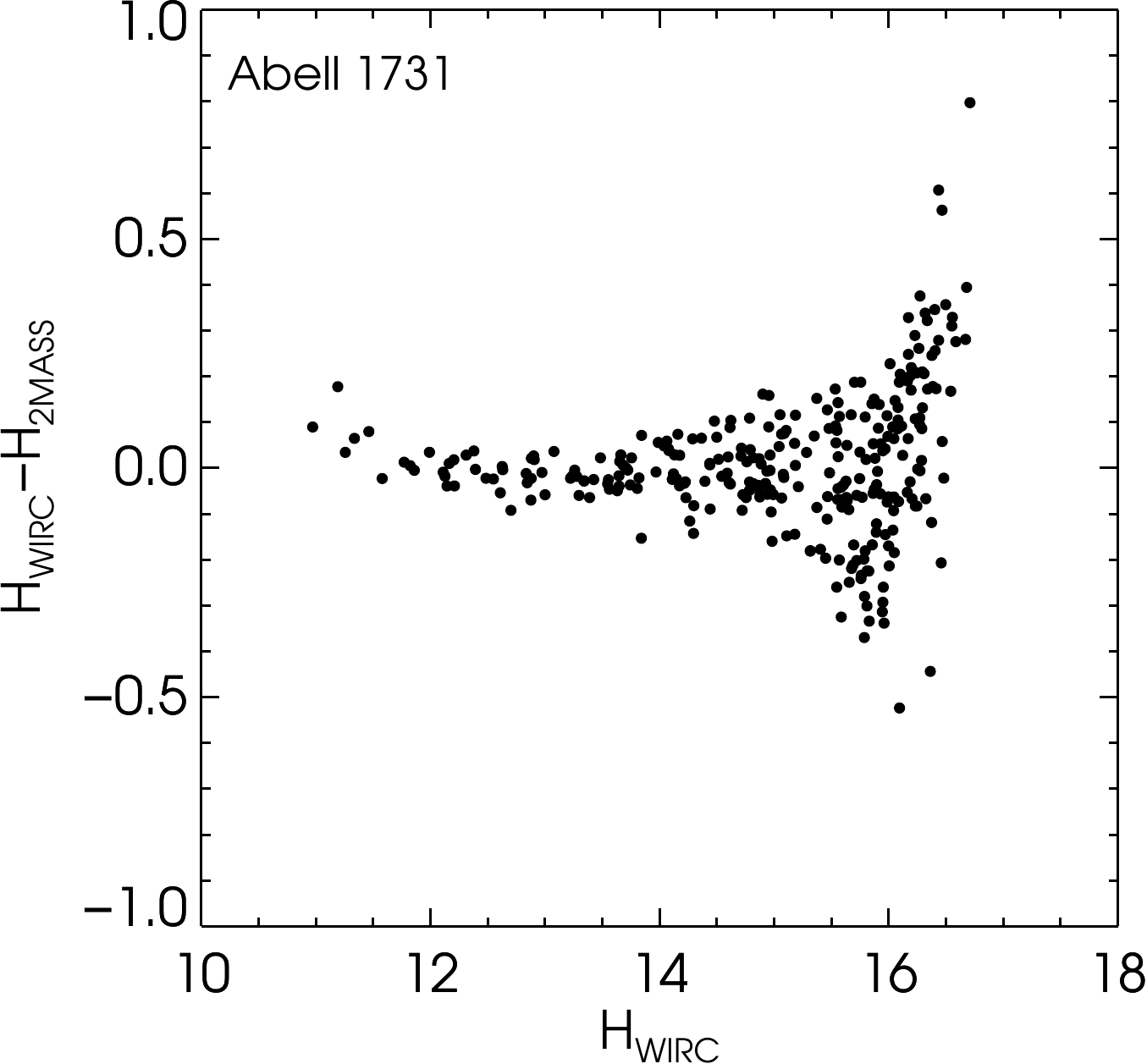}\includegraphics[width=0.3\linewidth, keepaspectratio]{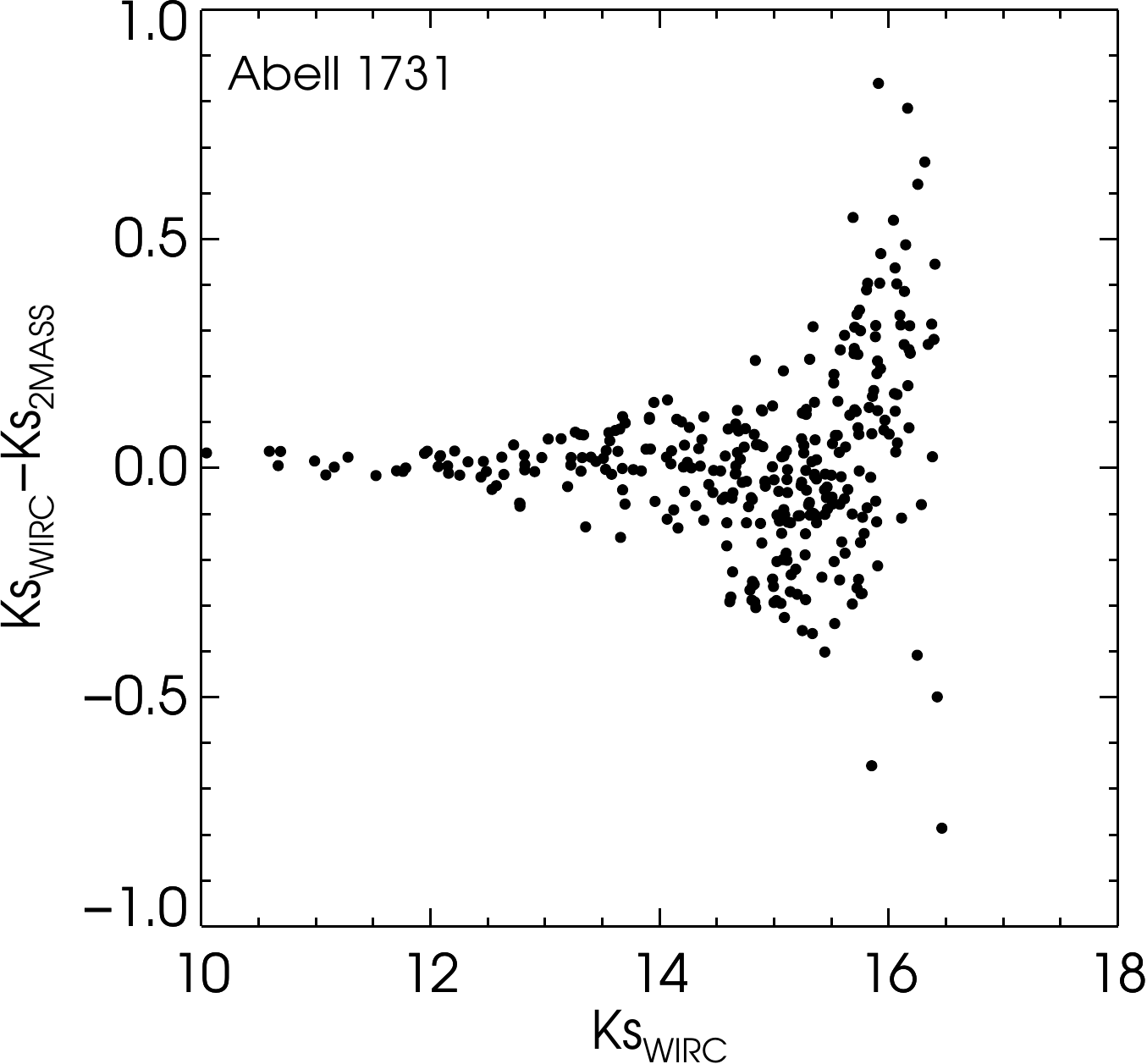}
\caption{The difference between the WIRC aperture magnitudes with the
  2MASS aperture magnitudes, plotted against the IRC aperture magnitudes, for A983 (top) and A1731 (bottom). Each panel
  presents the sources in our field of view with a 2MASS
  counterpart and after the removal of close pairs.}\label{2mass_comp}
 \end{figure*}

In order to compare our photometry with 2MASS, aperture magnitudes
were extracted from the mosaicked J, H and Ks images, using SExtractor
and a 4~arcsec aperture radius. This aperture was chosen to match the one
used in 2MASS. In Figure~\ref{2mass_comp}, we present the comparison
between our WIRC aperture magnitudes and the 2MASS ones. We measured a higher dispersion (1$\rm\sigma$) of the difference between the two magnitudes at the faint end but limited to about $\rm3.5\%$ of the corresponding WIRC magnitude. 
This value decreases towards higher magnitudes ($\rm<15$) to about $\rm2.5\%$. The
increasing scatter at the faint end is also due to the increase of the noise in the 2MASS data, that are close to the detection limit.

We then proceeded with the measurement of the Petrosian magnitudes
from our mosaicked images using SExtractor. These magnitude, used for the final catalogue, are measured on circular apertures
within Petrosian radii. The Petrosian radius is defined as the radius
at which the local surface brightness is a factor of 0.2 times the
mean surface surface brightness inside the radius \citep{petrosian76}.
In order to be able to detect both faint and extended sources, we set
a \textsl{det\_minarea} of $5$ and \textsl{det\_minthresh} of
$2.0$. We used a \textsl{deblend\_mincont} of $0.00005$ to detect the
smallest object that might be close to the largest galaxies.

 \begin{table}

\begin{center}
\begin{tabular}{lccc}
\hline
Channel & Aperture & Aperture & Color \\
 & radius ["]  & corr. & corr.\\
\hline
\hline
IRAC1 & 3 &1.14 & ... \\
IRAC2 & 3 &1.14 & ...\\
IRAC3 & 3 &1.25  & ...\\
IRAC4 & 3 &1.42  & ...\\
MIPS24 & 10 &1.167 & 1.041 \\
MIPS70 & 16 &2.044 & 1.089\\
MIPS160 & 20 &3.124 & 1.043\\
\hline
\end{tabular}
\end{center}
\caption{Aperture and color corrections applied to the IRAC and MIPS sources.}
\label{ap_corrections}
\end{table}
\subsection{Observations: optical spectra at WIYN}\label{sec_opt}

We obtained optical spectra of the MIPS $24\mu$m sources with flux greater than 0.3 mJy and optical magnitudes less than 20.5 in r' band in two
different runs using the Hydra instrument mounted at the WIYN
telescope at Kitt Peak National Observatory (KPNO). In total, we
observed six  and four overlapping pointings for A1731 and A983, respectively (see Figure~\ref{coverage} and Table~\ref{hydra_obs}).

During the first run on 2006 April 28-30,
  we obtained spectra of all 24 $\mu$m sources in A1731. We also
  observed three WIYN fields of A983 targeting 8$\mu$m sources since
  the MIPS observations were not yet available at the time of
  observation.  During the second run on 2008 Jan 14 we completed the
  observations of the 24$\mu$m sources in A983 which were not already
  observed as 8$\mu$m emitters by observing a further 69 sources.  To
obtain spectra between 4000 and 9000 \AA\, we used the red cable and
grating 316@7.0 at an angle of 21 degrees, centered at 6500~\AA, and
integrated for $3\times 1200$ s. This gives a dispersion of
2.6~\AA\ pixel$^{-1}$, a spectral coverage of 5400\AA, and a
resolution of 5.7\AA\. The package
  \texttt{DOHYDRA}\footnote{F. Valdes 1995, Guide to DOHYDRA,
    available at http://iraf.noao.edu/tutorials/dohydra/dohydra.html}
  was used for the extraction and reduction of the spectra including
  the wavelength calibration, application of the flat-field and fiber
  throughput correction, and the sky subtraction. Cosmic rays were
removed using the \texttt{la\_cosmic} task \citep{vandokkum01}. The
scripts were run for each configuration and for each night. More
details on the reduction methods can be found in \citet{marleau07}.
Considering the available 101 fibers, on average ten fibers were not usable and ten others were assigned to sky observation at each pointing. The number of 24 $\rm\mu$m target sources per configuration was less than the number of available fibers. Therefore, these free fibers were positioned on r'-band sources with no 24$\mu$m counterpart (on average 8 per configuration). We were able to obtain a total of 281 and 406 spectra for A983 and 1731, respectively. Of these, only 10 and 25 spectra, respectively,  did not have identifiable lines or continuum features.

We added archival SDSS DR10 spectra to our WIYN data. We selected the SDSS sources that are located in the cluster regions and extending up to $\sim8$ Mpc ($\rm\sim 4 \,r_{200}$) in clustercentric distance. Our aim was to increase the number
of sources with spectroscopic data at the outskirt of the cluster.  The line fitting was performed with an IDL code for
both the WIYN and SDSS spectra. The code includes the Markwardt
package algorithm for curve fitting. The region of the local continuum
is selected by hand and fit with a straight line. The line is fitted on top of the continuum with a Gaussian function. In case of blended lines, the code handles the fit of
multiple Gaussian functions. The measured flux corresponds to the set
of Gaussians that minimizes the $\chi^2$ value of the fitted
lines. The redshift of each line is allowed to vary and the final redshift assigned to the spectrum is the mean of the redshift of each line.  The absorption features 
 are included in the spectral templates. Due to the level of noise, the fit to the absorption features was possible for only 4\% of the total sample of spectra. Accounting for the absorption features helps in recovering the total line flux that is otherwise underestimated. This effect is more relevant at the high order of the 
Balmer series ($\rm H_{\delta}$, $\rm H_{\gamma}$), where the fraction of the absorbed flux to the emission line flux is higher. The $\rm H_{\alpha}$ and $ \rm H_{\beta}$ lines are dominated by the emission and the absorption accounts for 1\% of the flux.  In case of the $\rm H_{\alpha}$ line, this translates to an average SFR of $\rm 0.05 M_{\odot}\, yr^{-1}$. The emission line fluxes are used to estimate the star formation rate and their
ratio is a stringent diagnostics of the presence of AGN. The average flux density limit that we achieved was $\rm 10^{-17}\,erg \,cm^{-2}\,s^{-1}\,Hz^{-1}$.

We compared our line fluxes to the SDSS photometry.
For each WIYN fiber configuration, we measured
the r'-band magnitudes using the SDSS DR10 filter response function.
A two arcsec aperture was used, corresponding to the diameter of the SDSS fiber.
These factors were computed for each spectrum with a SDSS counterpart. The median value of this flux correction factor was then calculated and the flux of each spectrum, including those without a SDSS counterpart, was corrected accordingly. This  correction factor ranged from 0.78 to 1.22  for both clusters.
In addition, we applied an aperture correction  to each H$\alpha$ flux measurement. This aperture correction was obtained by taking the ratio of the fluxes extracted from the SDSS r'-band image using a 2 and a 10~arcsec diameter aperture. The latter aperture was chosen as it corresponds to the average size of our sources.

\begin{table}
\begin{center}
\begin{tabular}{lcl}
\hline
Column & Format & Description\\
\hline
\hline
1 & i6 & Catalog Number\\
2 & a12 &R.A. (J2000)\\
3 & a12 &Dec. (J2000)\\
4 &f10.3 &MIPS24 ap 10"($\mu$Jy)\\
5 &f10.3 &MIPS24 ap 10" error ($\mu$Jy)\\
6 &f10.2 &GALEX NUV ($\mu$Jy)\\
7 &f10.2 &GALEX NUV error($\mu$Jy)\\
8 &f10.3 &u$^{\prime}$ ($\mu$Jy)\\
9 &f10.3 &u$^{\prime}$ error ($\mu$Jy)\\
10 &f10.3 &g$^{\prime}$ ($\mu$Jy)\\
11 &f10.3 &g$^{\prime}$ error ($\mu$Jy)\\
11 &f10.3 &r$^{\prime}$ ($\mu$Jy)\\
13 &f10.3 &r$^{\prime}$ error ($\mu$Jy)\\
14 & f10.3 &i$^{\prime}$ ($\mu$Jy)\\
15 & f10.3 &i$^{\prime}$ error ($\mu$Jy)\\
16 &f10.3 &z$^{\prime}$ ($\mu$Jy)\\
17 &f10.3 &z$^{\prime}$ error ($\mu$Jy)\\
18 & f10.3 &J Petrosian ($\mu$Jy)\\
19 & f10.3 &J Petrosian error ($\mu$Jy)\\
20 & f10.3 &H Petrosian ($\mu$Jy)\\
21 & f10.3 &H Petrosian error ($\mu$Jy)\\
22 & f10.3 &K$_{s}$ Petrosian ($\mu$Jy)\\
23 & f10.3 &K$_{s}$ Petrosian error ($\mu$Jy)\\
24 & f10.2 &IRAC1 ap 3" ($\mu$Jy)\\
25 & f10.2 &IRAC1 ap 3" error ($\mu$Jy)\\
26 & f10.2 &WISE1  ($\mu$Jy)\\
27 & f10.2 &WISE1 error ($\mu$Jy)\\
28 & f10.2 &IRAC2 ap 3" ($\mu$Jy)\\
29 & f10.2 &IRAC2 ap 3" error ($\mu$Jy)\\
30 & f10.2 &WISE2  ($\mu$Jy)\\
31 & f10.2 &WISE2 error ($\mu$Jy)\\
32 & f10.2 &IRAC3 ap 3" ($\mu$Jy)\\
33 & f10.2 &IRAC3 ap 3" error ($\mu$Jy)\\
34 & f10.2 &IRAC4 ap 3" ($\mu$Jy)\\
35 & f10.2 &IRAC4 ap 3" error ($\mu$Jy)\\
36 &f10.2 &MIPS70 ap 16" ($\mu$Jy)\\
37 &f10.2 &MIPS70 ap 16" error ($\mu$Jy)\\
38 &f10.2 &MIPS160 ap 20" ($\mu$Jy)\\
39 &f10.2 &MIPS160 ap 20" error ($\mu$Jy)\\
40 &f9.3 & MIPS24 SNR in 6" aperture\\
\hline
\end{tabular}
\end{center}
\caption{The IR Source Catalog columns.}\label{catalogue_entries}
\end{table}

\subsection{Archival data photometry}

We obtained archival data to complement the wavelength
coverage of our observations.  We retrieved the five optical bands
\textit{u' g' r' i' z'} from SDSS DR10, that are needed to measure a stellar mass estimate for each galaxy, using spectral energy
distribution (SED) fitting. We also included the near UV (NUV) band
from GALEX, containing the emission from newly formed stars.
Additionally, we retrieved the archival WISE 3.6 and $4.6\mu$m
data. These bands were used to check the photometry of the IRAC 1 and 2 channels. The 1$\rm\sigma$ dispersion of the difference, with respect to the IRAC magnitude, between the IRAC and WISE photometry ranged from $\rm5\%$ to $\rm3.5\%$ from low to high magnitudes, respectively.

\begin{figure*}
 \centering
  \includegraphics[width=0.5\linewidth, keepaspectratio]{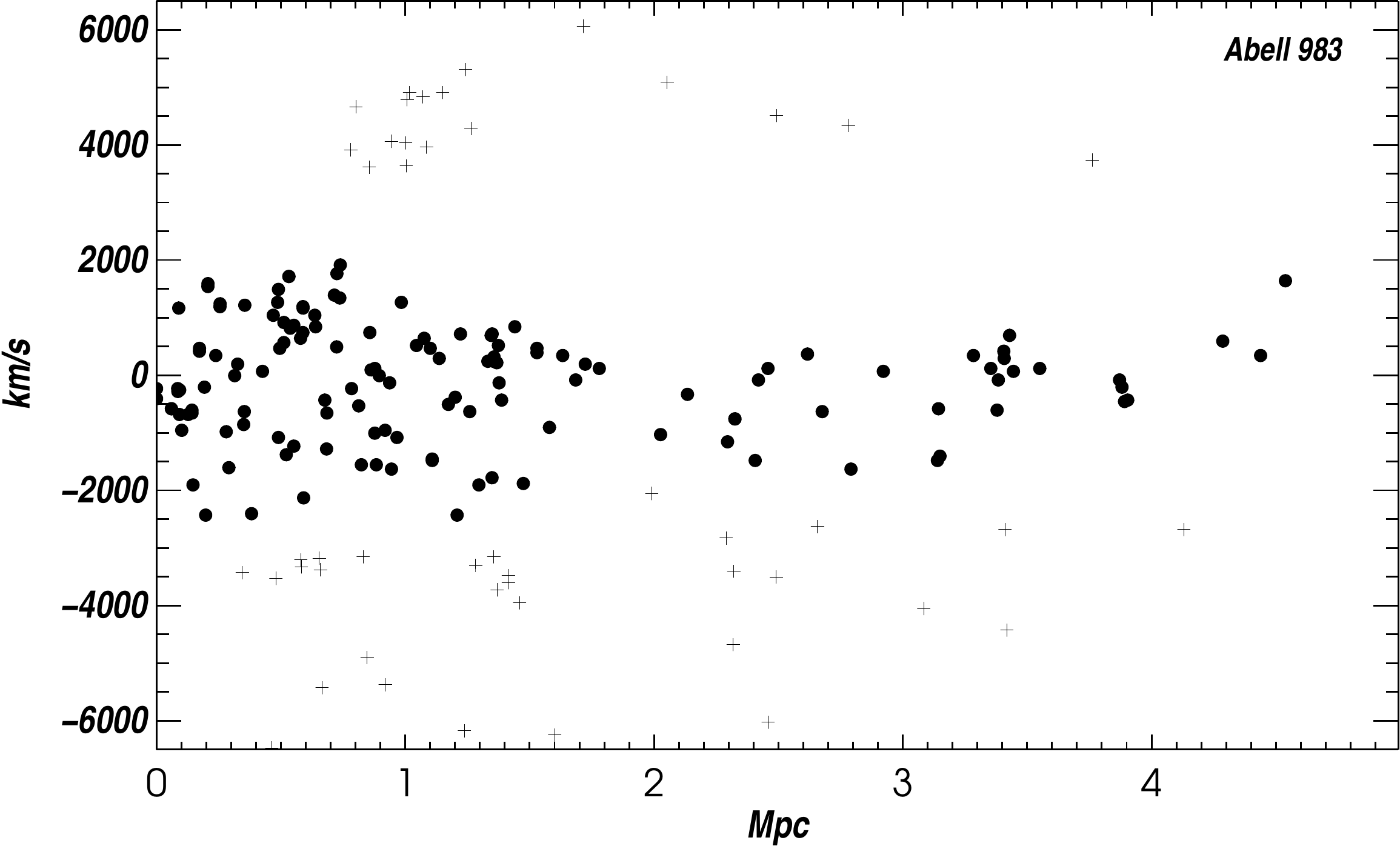}\includegraphics[width=0.5\linewidth, keepaspectratio]{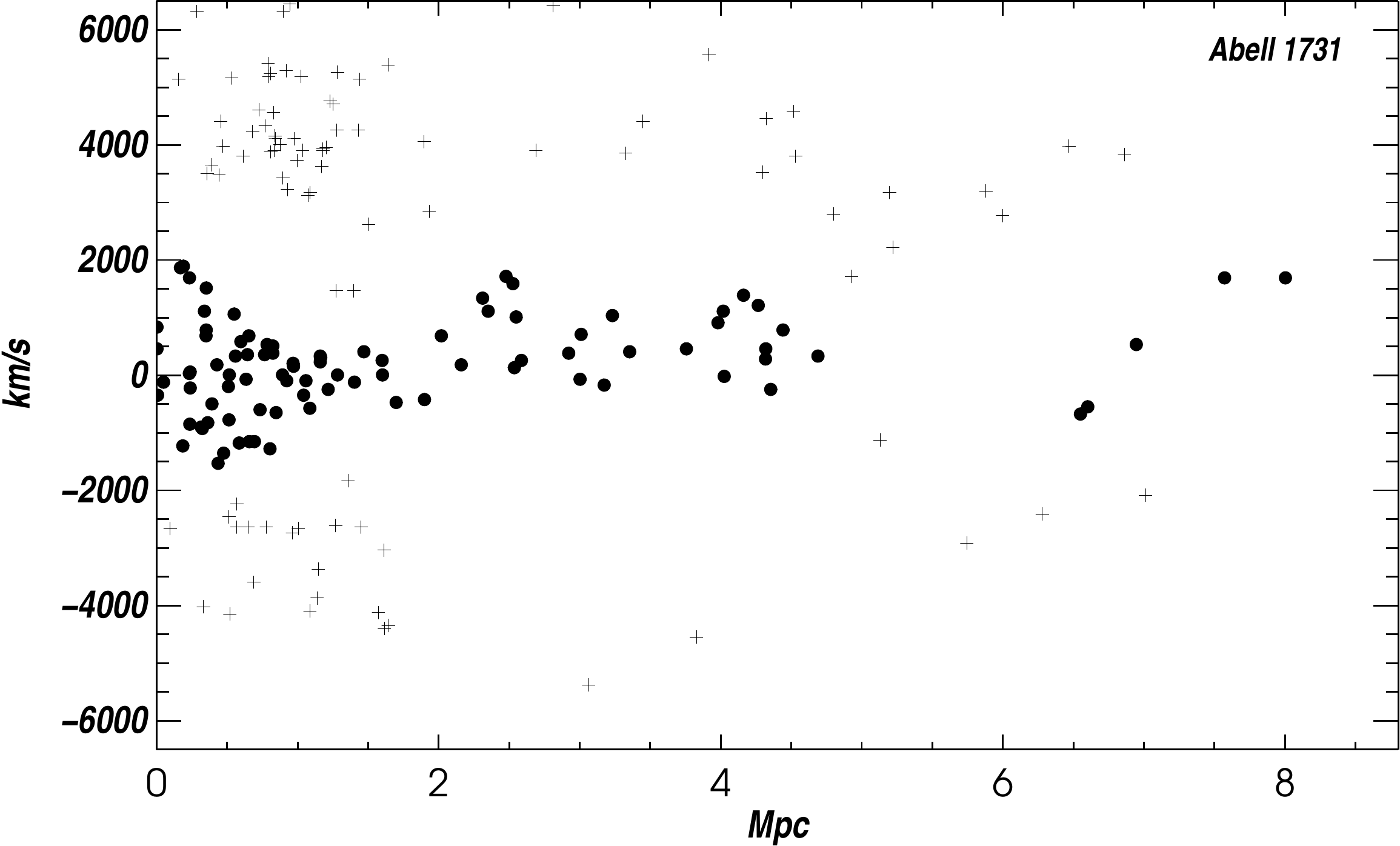}
\caption{The observed galaxies with a spectroscopic redshift in the
  range $0.15<z<0.25$, plotted in the clustercentric distance -- velocity
  space. The filled dots correspond to the cluster members selected
  with the shifting gap method by \citet{fadda96}. The right and left
  panels correspond to A983 and 1731,
  respectively.}\label{members}
 \end{figure*}

\section{Cross matching of the source catalogues}\label{cat}

We produced a final photometric catalogue using as reference the positions of the MIPS
$\rm 24\mu$m sources with a SNR in the first Airy ring ($\sim$
6~arcsec) larger than 3. The matching distance used corresponds to half of
the first Airy ring (3~arcsec). The matching algorithm associates
the position of the $\rm 24\mu$m sources to sources detected in the NUV, the five SDSS optical bands, the three near-IR bands, the 
four mid-IR bands, the other two far-IR bands and the spectroscopic redshift.
The spatial distribution of the sources is sparse in the field. Nonetheless, the FWHM of the MIPS $\rm 24\mu$m is
sufficiently large to hide close pairs of sources. A comparison with the IRAC images, which have a higher resolution, revealed that this
problem affected only a small fraction of the total number of sources
($\rm \sim 1\%$). The spectroscopically confirmed members  (see Section~\ref{sec_members}) of the two clusters were not affected by this issue, since all of them have a unique counterpart in all the wavebands.  Table~\ref{catalogue_entries} summarises the entries of our
IR selected source catalogue.

\section{Data analysis}\label{data_analysis}

\subsection{The cluster membership}\label{sec_members}

The
peculiar velocity of each galaxy  was computed with respect to the mean cluster velocity $\rm \bar{v}=\bar{z}c$, where $\bar{z}$ is the redshift of the cluster and c is the speed of light. The redshift of the cluster was computed as the mean value of the redshifts in the range $0.19<$z$<0.21$.  We used the shifting gap algorithm, as described in 
\citet{fadda96} to determine the cluster membership of the galaxies we observed. The
shifting gap method makes use of both galaxy velocities and
clustercentric distances.  The clustercentric distance to each cluster galaxy was measured from the brightest cluster galaxy (BCG). The galaxies were grouped in overlapping and shifting bins of $\rm 500\,\rm kpc $ from the cluster center (or wide enough to
contain at least 15 or 20 galaxies each for A983 and 1731, respectively). Then, gaps of $\rm 1000 \,km\,
s^{-1}$ and $\rm 800 \,km\,
s^{-1}$ for A983 and 1731 respectively, were searched in the observed galaxy velocity distribution. These gaps mark the separation between the velocity distribution of the cluster members and the external galaxies. This procedure rejects
galaxies (interlopers) that have velocities bigger than $\rm 1000 \,km\,
s^{-1}$ and $\rm 800 \,km\,
s^{-1}$, for A983 and A1731, respectively. The procedure was iterated until the number of cluster
members converged.  The advantage of this
statistical method is the independence from any physical assumption on
the dynamical state of cluster. The shifting gap method gives us a
total of 134 and 91 members for A983 and 1731, respectively (see Figure~\ref{members}). The procedure was run iteratively
until convergence on the number of selected members. As a
comparison, we run also the algorithm of \citet{mamon10}. This code
evaluates cluster membership from the distance-velocity space, based
on models of mass and velocity anisotropy of cluster haloes obtained
from a cosmological simulation.  This method produced
us 105 members and 77 members for A983 and 1731. This code outputs estimates for the virial mass, the velocity dispersion and $\rm r_{200}$, and are quoted in Table~\ref{cluster_props}. The bias towards late-type galaxies of our sample could lead to an overestimate of the velocity dispersion, as these objects present more elongated orbits with respect to the passive population of galaxies \citep{biviano04}. All the members selected via the \citet{mamon10} method are included in the sample selected via the shifting gap algorithm. As the shifting gap method appears to be more conservative in identifying cluster members, specifically at high clustercentric distances, we keep this
larger sample for our analysis.
\begin{figure}
 \centering
  \includegraphics[width=0.49\linewidth, keepaspectratio]{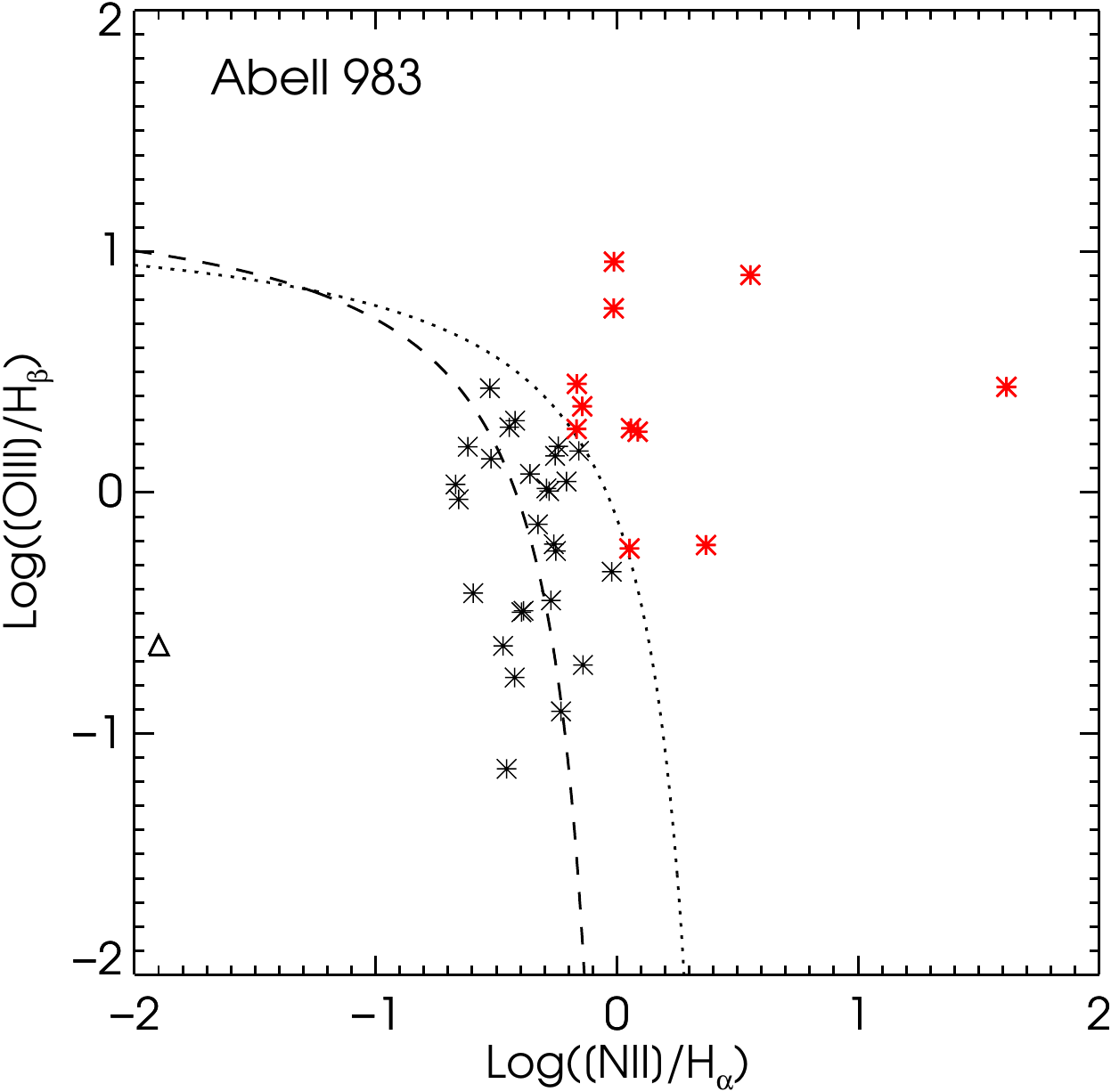} 
  \includegraphics[width=0.49\linewidth, keepaspectratio]{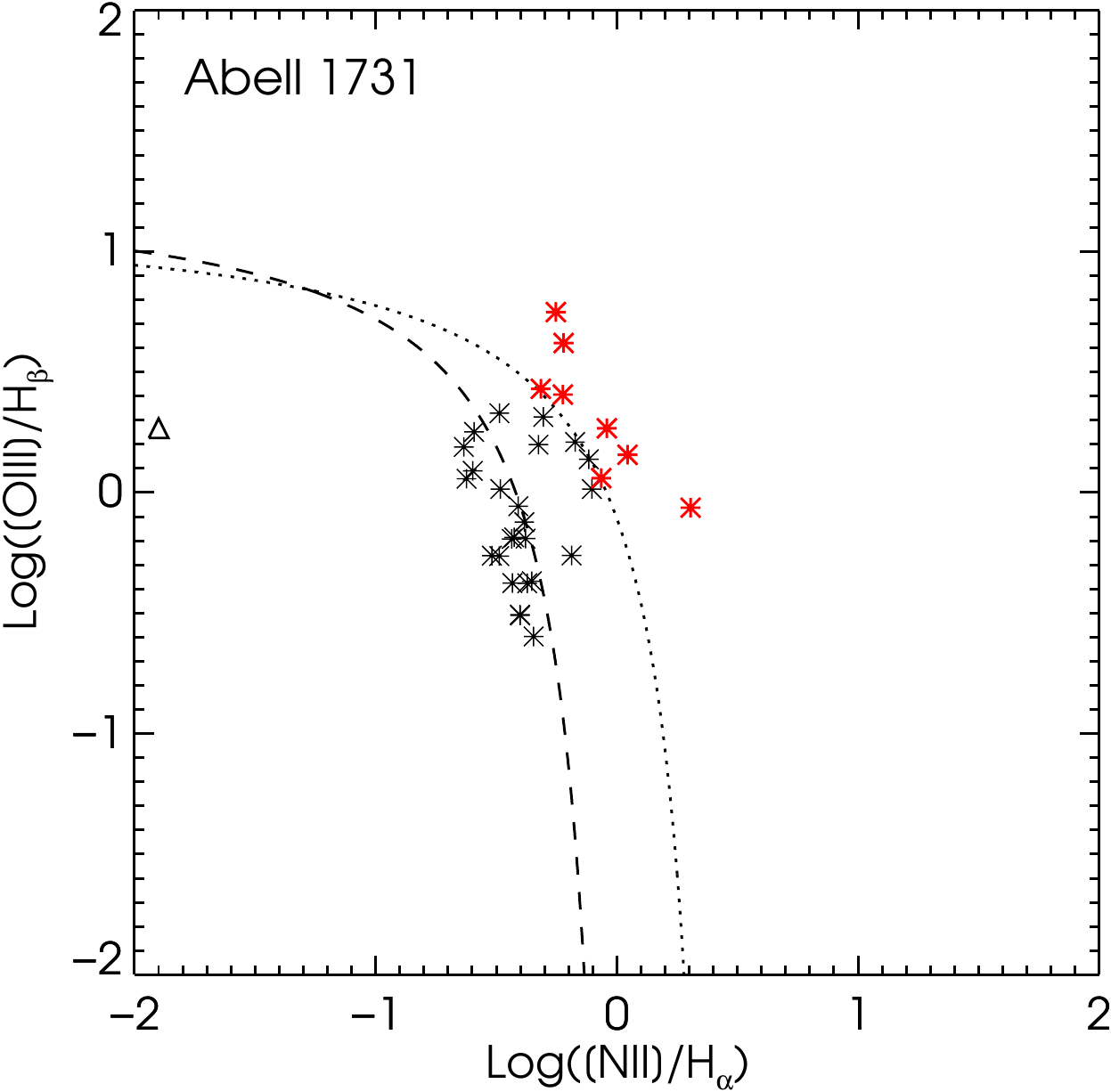} 
\caption{Emission line diagnostic diagrams using $\rm[NII]/H_{\alpha}$
  and $\rm[OIII]/H_{\beta}$ line ratios for A983 and A1731 in the left and right panel, respectively. The
  dotted line separates the HII star forming regions, below, from the
  AGN, above, following the modelling by \citet{kewley01}. The selected AGN candidates are plotted as red asterisks. The
  triangle identifies the source without and $\rm[NII]$ emission
  lines. As a
  comparison, we plotted also the \citet{kauffmann03} cut as dashed
  line, that selects less fiercely star forming
  galaxies.}\label{bpt}
  \vspace{0.2cm}
 \hspace{-0.3cm}\includegraphics[width=0.51\linewidth, keepaspectratio]{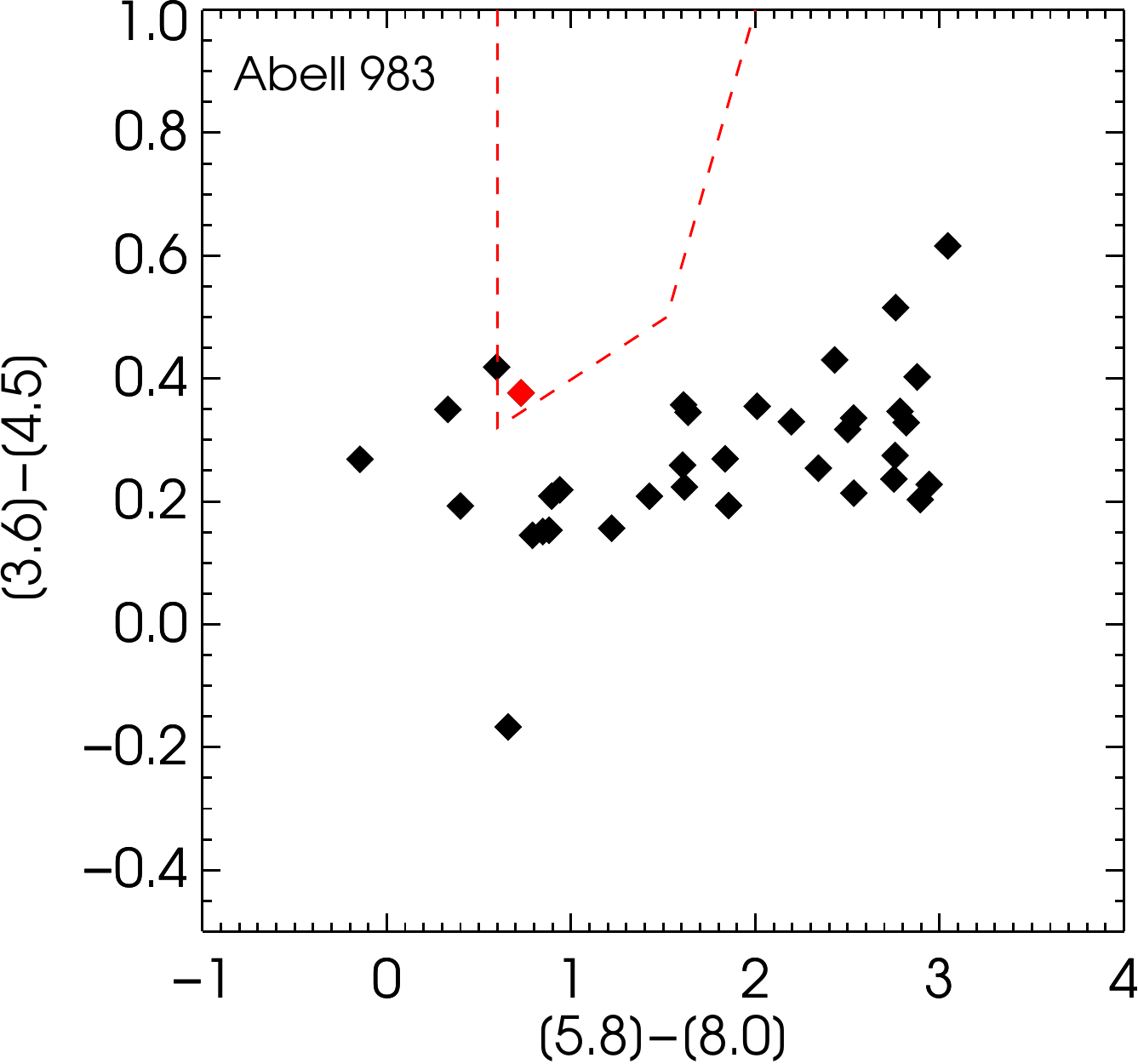}\includegraphics[width=0.51\linewidth, keepaspectratio]{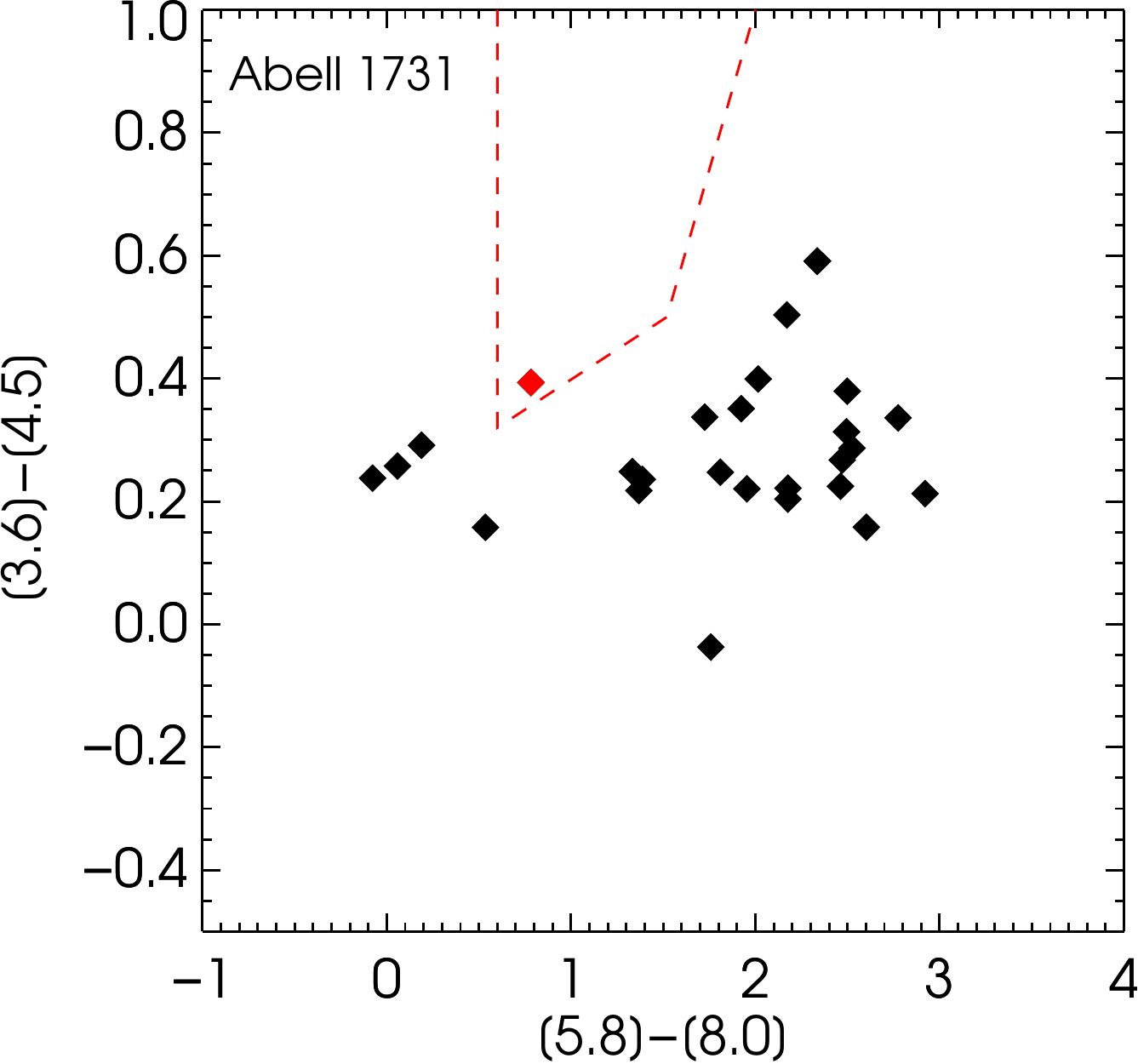}
\caption{AGN selected with the \citet{stern05} color selection
  criteria. Each color was evaluated by converting the flux in each of
  the IRAC channels into Vega magnitudes. Galaxies in A983
  and 1731 are plotted in the left and right panel,
  respectively.}\label{agn_stern}
 \end{figure}

\subsection{Stellar masses}\label{sec_stellar_mass}
We estimated the stellar masses for each cluster member from the SED
fitting, using the software MAGPHYS \citep{dacunha08}, with the
addition of the stellar libraries by \citet{bruzual03}.  The
wavelength coverage (up to 16 bands) and the precision of the
spectroscopic redshift ensure the high significance of the fit. As described in Section~\ref{sec_opt}, the spectroscopic observations targeted also non IR emitters. For these
galaxies, we computed a separate SED fit that included only the bands
from NUV to Ks.

\subsection{AGN and star forming galaxy separation}

The SFR values were retrieved using two different methods, i.e. using
the computed luminosity in the infrared bands and the $H_{\alpha}$
emission line flux. This two methods assume that the emission is dominated by star formation. Hence,  before we applied these methods, we identified those objects whose emission was dominated by an AGN.  We used three independent
diagnostics for the detection of the AGN: the first is based on the
characteristic ratio of  optical emission lines, the second on identifying broad line AGN and the third utilises an infrared color diagnostic.

\subsubsection{The emission line diagnostic}\label{agn_el}

The ratio of the fluxes of specific emission lines is useful for
discerning the source of the ionizing radiation causing such lines.  A
clear signature of an active galactic nucleus is a high value for the
flux ratio of $\rm[NII]/H_{\alpha}$ and $\rm[OIII]/H_{\beta}$
\citep{baldwin81}, with respect to more moderate values in case of a
star forming region. In order to identify narrow-line AGN in our
spectroscopic sample, we used the diagnostic diagrams of
$\rm[NII]/H_{\alpha}$ versus $\rm[OIII]/H_{\beta}$ and $\rm[OIII]/H_{\beta}$
versus $\rm[SII]/H_{\alpha}$, applying the selection criterion by
\citet{kewley01}. These cuts result from the modelling of starburst
galaxies with stellar population models (PEGASE version 2.0),
producing the ionizing radiation, and with a detailed self-consistent
photoionization model (MAPPINGS III). The AGN are modeled  as $500\rm \,km \,s^{-1}$ radiative shocks \citep{kewley01}. For A983, we identified 11
and 6 AGN using the two diagrams with $\rm[NII]$ and $\rm[SII]$,
respectively. Only two candidate AGN are common to both diagrams. In the case of A1731, we
identified 8 and 7 AGN using both $\rm[NII]$ and $\rm[SII]$ diagrams, respectively
(see Figure~\ref{bpt}). In this case, only one candidate AGN is common to both diagrams. We consider only the AGN selected via the $\rm[NII]$ as the $\rm[SII]$ lines (with observed wavelength $\sim8000$\AA)  are located towards the end of the observed waveband range($\sim9000$\AA), where the sky emission lines dominate. Furthermore, \citet{kewley06} showed that the $\rm[NII]$ selection method is more sensitive to low energy AGN. In addition to these narrow line emission
AGN, we visually identify 5 and 4 broad emission line AGN in A983 and 1731,
respectively. Broad emission line AGN present permitted line width in the range $\rm 10^3-10^4\,km\, s^{-1}$, while narrow emission line AGN show permitted and forbidden line width of $\rm 10^2-5\times 10^2\,km\,s^{-1}$  \citep{hao05}.

 \begin{figure}
 \centering
   \includegraphics[width=0.5\linewidth, keepaspectratio]{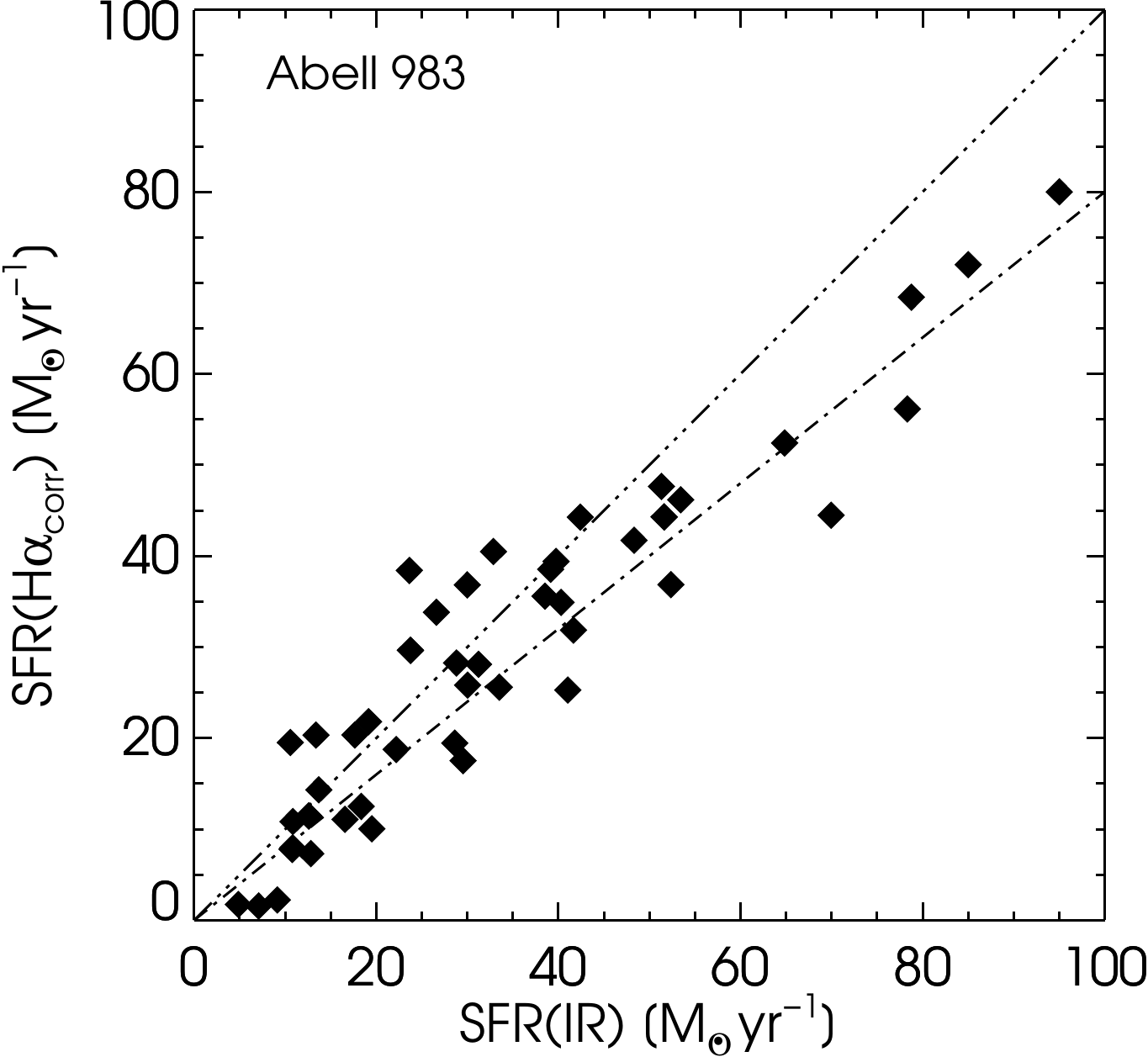}\includegraphics[width=0.5\linewidth, keepaspectratio]{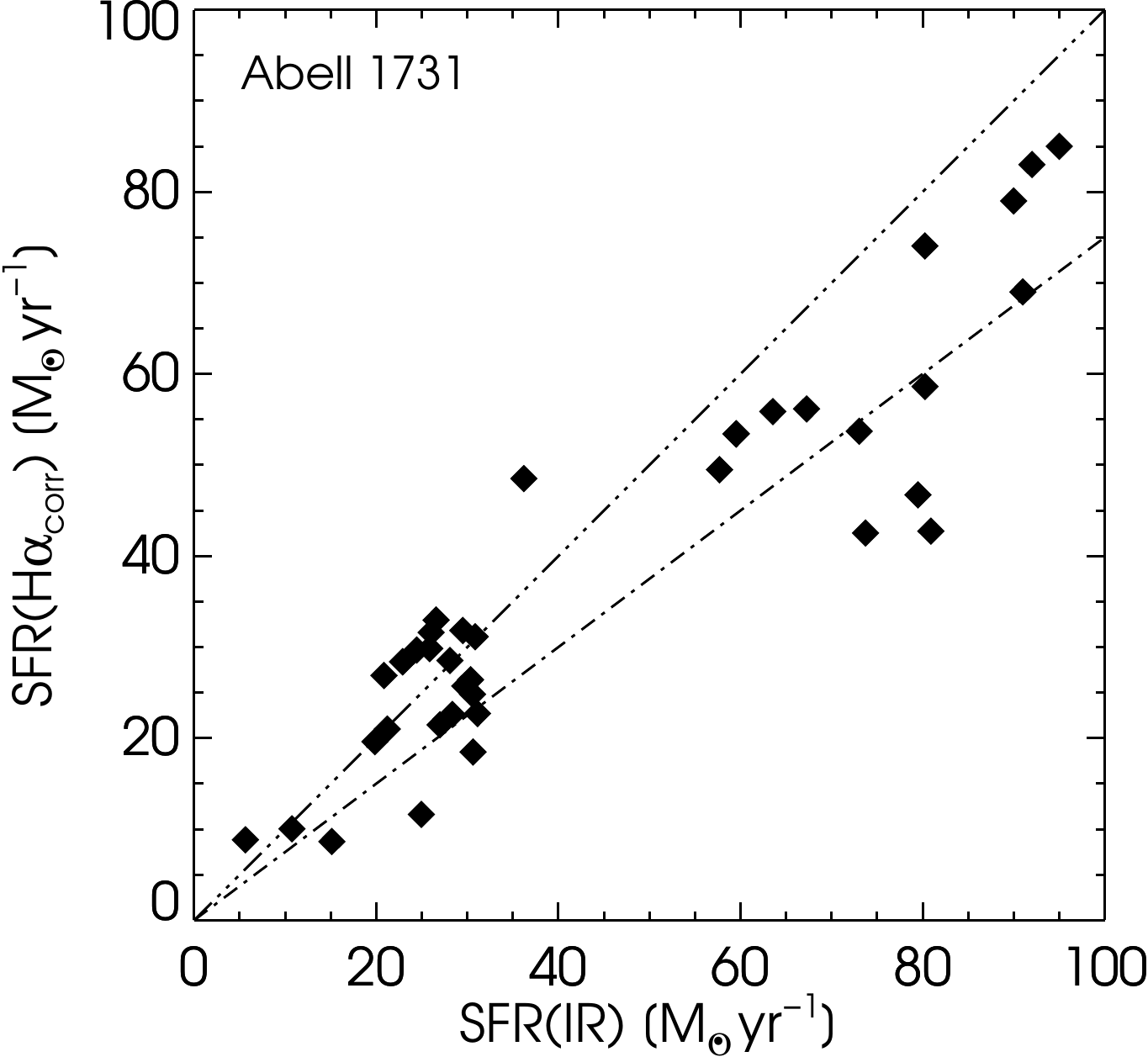}
  \caption{The comparision of SFR evaluated from the extinction corrected $\rm H\alpha$ line flux and the SFR based on the total IR luminosity obtained from the SED fit. The triple dot dash line corresponds to the relation with slope m=1. The dot dashed line corresponds to the linear fit of the data. ($\rm m=0.8\pm0.02$ and $\rm m=0.74\pm0.06$ for A983 and A1731, respectively). The left and right panels refer to A983 and 1731, respectively.}\label{sfr_ext}
   \end{figure}

\subsubsection{The IR color diagnostic}\label{agn_IR}   

We used the IR color selection proposed by \citet{stern05} to identify obscured
AGN in our sample. The sources within the so-called "Stern
  wedge" are tagged as AGN. For each cluster, we find only one AGN
candidate (see Figure~\ref{agn_stern}), not identified  using the emission line diagnostic. In total we identify 17  AGN candidate in A983 and 13 in A1731.

\subsection{The SFR from the total IR luminosity}\label{sec_sfr_ir}

The use of MAGPHYS coupled with our multiwavelength catalogue
(up to 16 bands from near-UV to mid-IR) ensures a consistent modelling
of the galaxies, that includes the different phases of the ISM and the
reprocessed star formation emission. We used the Kennicutt relationship
\citep{kennicutt98} to translate the total infrared luminosity $\rm L_{IR}$ (from 8
to 1000$\mu$m) estimated by MAGPHYS into a SFR:
\begin{equation}
\rm SFR[M_{\odot}\, yr^{-1}] = 1.7 \times 10^{-10}\, L_{IR}/L_{\odot}.
\end{equation}

 \begin{figure*}
   \centering
   \includegraphics[width=0.5\linewidth,keepaspectratio]{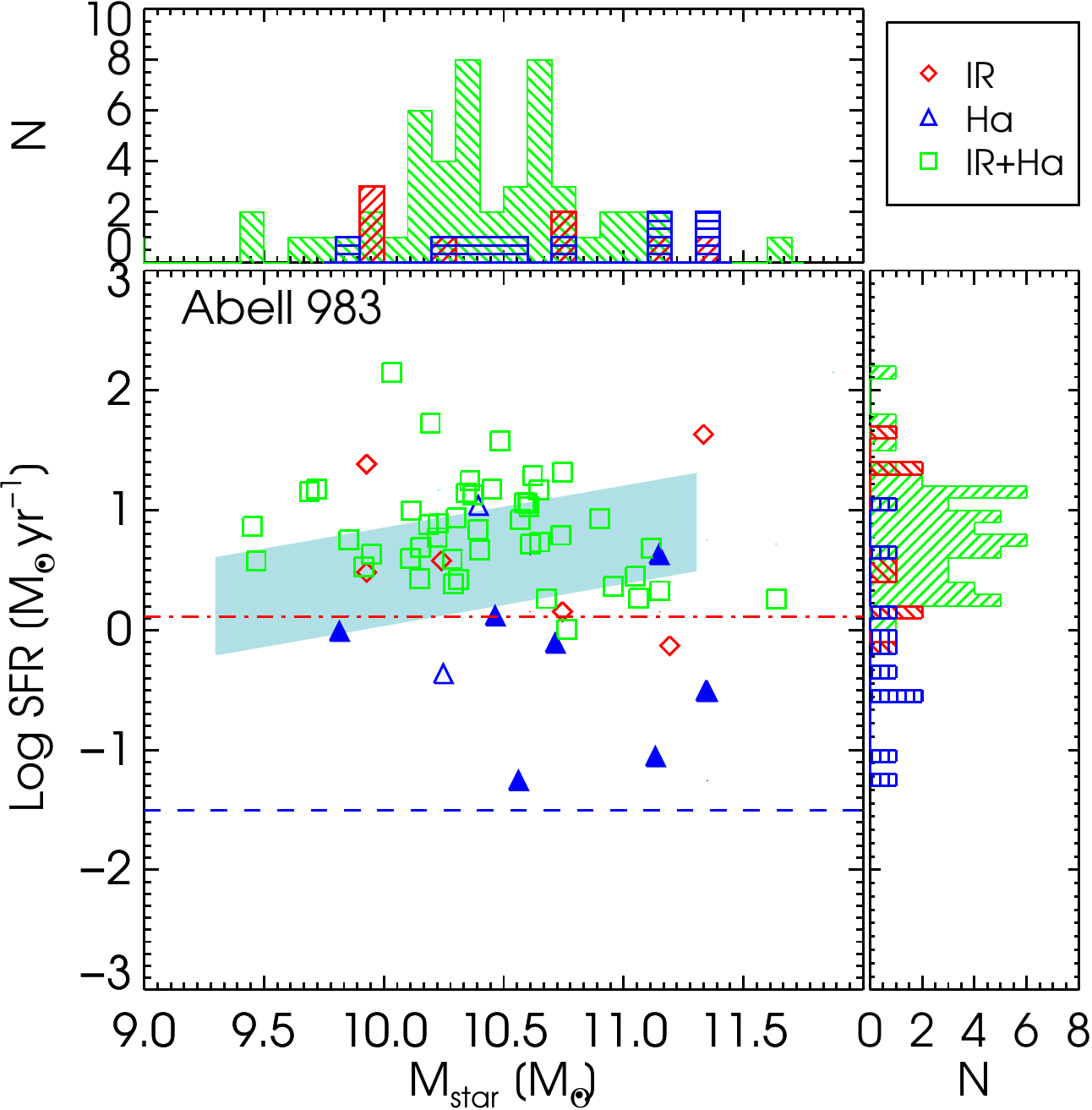}\includegraphics[width=0.5\linewidth,keepaspectratio]{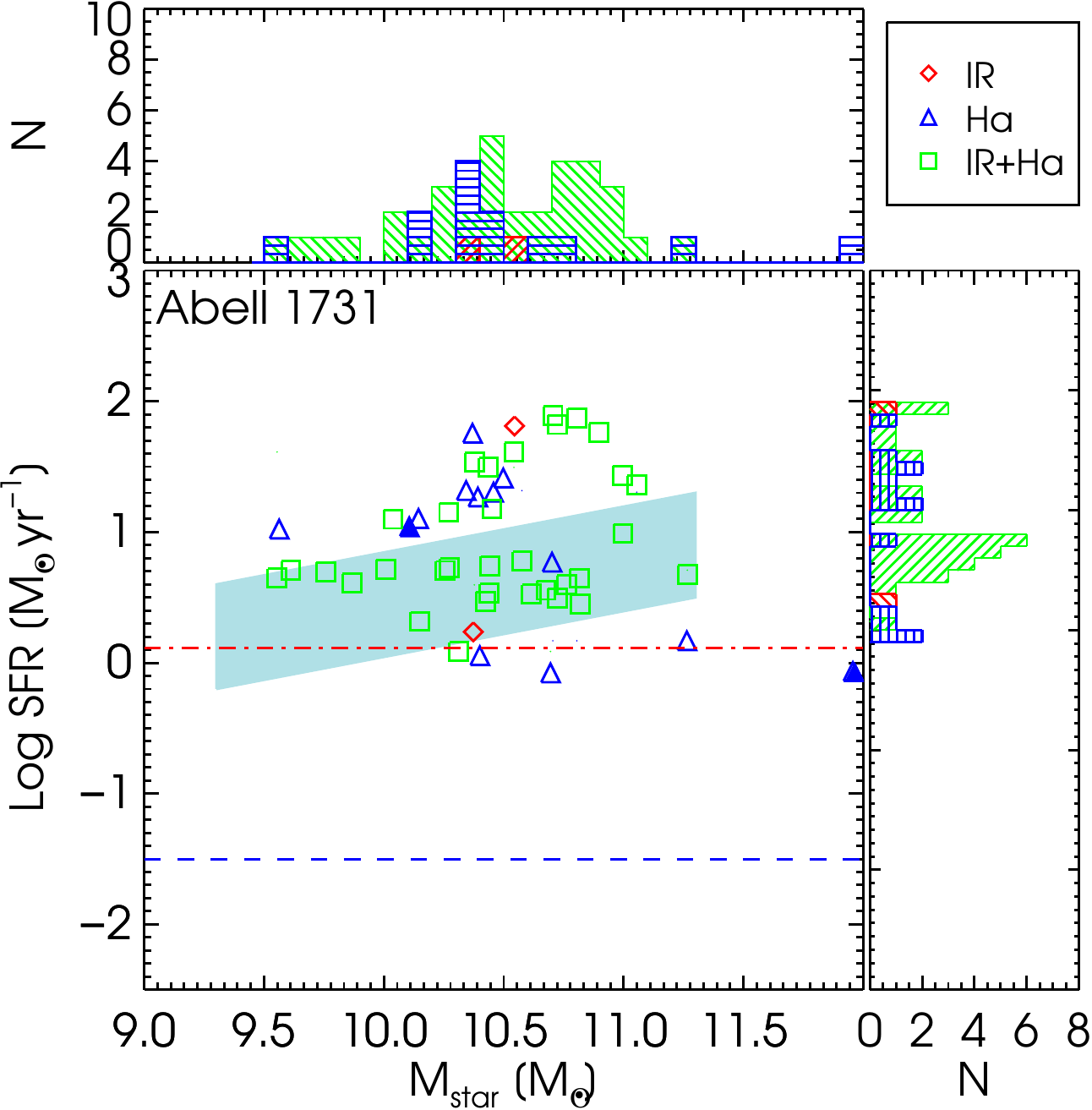}
 \caption{The $M_*-SFR$ relation for the A983 and 1731 on the
   left and right panel, respectively. The histograms on the top and right side of each plot show the stellar mass and SFR distribution of the members. The SFR from the total IR
   luminosity only is plotted as red diamonds, the SFR from the
   extinction corrected $\rm H\alpha$ as open blue triangles, the combined SFR as
   open green squares. The filled blue triangles mark the sources for
   which the mean extinction correction was used. The light blue shaded area corresponds to the relation with uncertainties found by \citet{noeske07}. The red dot-dashed
   and the blue dashed lines mark the SFR limit evaluated from the
   detection limit of the IR $24\mu$m and optical spectral SNR,
   respectively.}\label{sfr_comb}
   \end{figure*}

\subsection{The SFR from the optical spectra}\label{sec_sfr_opt}

\subsubsection{The extinction correction}

The emission line fluxes were corrected for the internal
absorption of each galaxy due to the ISM.  The extinction can be
estimated using the Balmer decrement, i.e. comparing the observed and
predicted Balmer line fluxes ($\rm H\alpha$ at $\rm6563$\AA  and $\rm
H\beta$ at $\rm4861$\AA). A direct measurement of this  decrement was possible for 53 and 66 galaxies out of 108 and 223 in the redshift range 0.15<z<0.25 in A983 and A1731 field, respectively. We used the median value of the Balmer
decrement of these subsets for the remaining galaxies in which one of
the two lines was not detected. The color excess E(B-V) of each source
is computed by comparing the ratio of the observed lines
$\rm H_{\alpha}$ and $\rm H_{\beta}$, $\rm
F_o^{H_{\alpha,\beta}}$, with the predicted unobscured value
via the equation:

\begin{equation}
\rm E(B-V)=\frac{2.5}{k(H_{\alpha})-k(H_{\beta})}\log\left(\frac{F_o^{H_{\alpha}}/F_o^{H_{\beta}}}{F_i^{H_{\alpha}}/F_i^{H_{\beta}}}\right)
\end{equation}
where $\rm
F_i^{H_{\alpha,\beta}}$ is  the intrinsic unobscured flux of $\rm H_{\alpha}$ and $\rm H_{\beta}$, respectively,  and $\rm k(\lambda)$ is the reddening curve as a function of the wavelength. Here, the intrinsic unobscured line ratio $ \rm
F_i^{H_{\alpha}}/F_i^{H_{\beta}}$ is set equal to 2.87, assuming
case~B recombination and $\rm T=10^4 \,K$ \citep{osterbrock89}. The
reddening curve $\rm k(\lambda)$ was taken from \citet{calzetti00}
for starburst galaxies in the wavelength range from $\rm 0.12-2.2
\mu$m. These quantities allow us to express the extinction as a
function of wavelength via:

\begin{equation}
\rm A(\lambda)= E(B-V)k(\lambda),
\end{equation}
where $A(\lambda)$ is the mean extinction in units of magnitude at a
specific wavelength $\rm \lambda$. Using the IDL routine
\texttt{calz\_unred}, we computed the dereddened flux for each
emission line. We assumed the default value for the effective total
obscuration for starburst galaxies $\rm R_{V}=4.05$, which include the
effect of extinction, scattering, and the geometrical distribution of
the dust relative to the emitters \citep{calzetti00}. The value of $\rm A(\lambda)$ ranges between 0.5 and 3.5 and its mean is 1.5.

\subsection{The star formation rate from the H$\alpha$ emission line}\label{sec_sfr_ha}

We calculated the SFR from the $\rm H_{\alpha}$ line flux (corrected for
aperture and extinction) using the $\rm H_{\alpha}$-SFR relation
derived by \citet{kennicutt98}:

\begin{equation}
\rm SFR(H\alpha) [\rm M_{\odot}\,
yr^{-1}]=7.9\times10^{-42} L(H\alpha) [erg \,s^{-1}]. \label{kenn_ha}
\end{equation}
This relation applies especially when considering young and
massive stellar populations, under the assumption that the emission
lines are tracers of the ionizing flux from newly formed
stars. Eq.~\ref{kenn_ha} was introduced and calibrated
using a Salpeter initial mass function (IMF) and over the mass range
$0.1 < M/M_{\odot} < 100$. We estimated star formation rates in the
range $\rm 0.1-200\, M_{\odot}\, yr^{-1}$. In agreement with \citet{marleau07}, we find that these SFR are strongly correlated with extinction.

\section{Results}\label{results}
\subsection{SFR from IR vs $\rm H_{\alpha}$}

Figure~\ref{sfr_ext} shows the comparison of the SFR obtained from
the IR luminosity and from the extinction corrected $\rm H_{\alpha}$ line flux.  An underestimate of the corrected flux, and hence of the SFR computed using the $\rm H_{\alpha}$, can be seen for galaxies with $\rm SFR > 30 \,\rm M_{\odot} \, yr^{-1}$. As a consequence, the linear fit of the data presents a slope of $\rm m=0.8\pm0.02$ and $\rm m=0.74\pm0.06$ for A983 and A1731, respectively.
The extinction correction is more effective for normal galaxies than for high star forming dusty ones \citep{calzetti00}. In order to properly account for the obscured and unobscured star formation, we added
the SFR from the total IR luminosity to the uncorrected $H_{\alpha}$ emission, following \citet{kennicutt09}. In the limit of complete obscuration (satisfied in the most active galaxies,  luminous and ultra-luminous infrared galaxies), 
the SFR was evaluated using the total IR luminosity only. The SFR of the
galaxies with optical spectral lines only was retrieved from the
corrected $H_{\alpha}$ emission.

Figure~\ref{sfr_comb} presents the $\rm M_*-SFR$ relation for the
spectroscopically confirmed members, along with 
the $\rm M_*$ and SFR distributions for each of the two clusters. We overplotted as a
comparison the relation obtained by \citet{noeske07} who used a sample of
field galaxies at reshift of 0.2$<$z$<$0.45. The different colours
encode the methods used for estimating the SFR. 
For both clusters, the mean SFR of the members is compatible with the mean SFR of a coeval sample of field objects ($\rm Log(SFR)\sim0.8\,M_{\odot} yr^{-1}$ for the clusters with respect to  $\rm Log(sSFR)\sim0.5M_{\odot} yr^{-1}$ for the field).

\subsection{The effect of the dynamical state of the cluster on SFR}\label{sec_state_sfr}

The clusters  that we are considering present a clear difference in their dynamical state. This difference is evident in the comparison of the clustercentric-velocity plot (Figure~\ref{members}) of the two
clusters.  A983 presents the classical trumpet-like shape of the
galaxy distribution, with a clear separation between the hosted
objects and the external ones. A1731 shows a less uniform distribution of the members. The central high number density of members becomes more sparse  at  a clustercentric distance of about $\sim 2\rm\,Mpc$ $\rm\sim 1\, r_{200}$). The analysis of the peculiar velocity of A1731 galaxy members allows us to exclude the presence of neighboring clusters or recent mergers, as no clear subclumps of galaxies are present. \citet{haines15} showed that  the accreting objects are supposed to be found at these characteristic radii, comparing the cluster in the LoCuSS survey and the 75 more massive clusters in the Millennium simulation \citep{springel05}. This might suggest that A1731
is still actively accreting galaxies from the field. 

\begin{figure}
  \centering
  \includegraphics[width=0.75\linewidth, keepaspectratio]{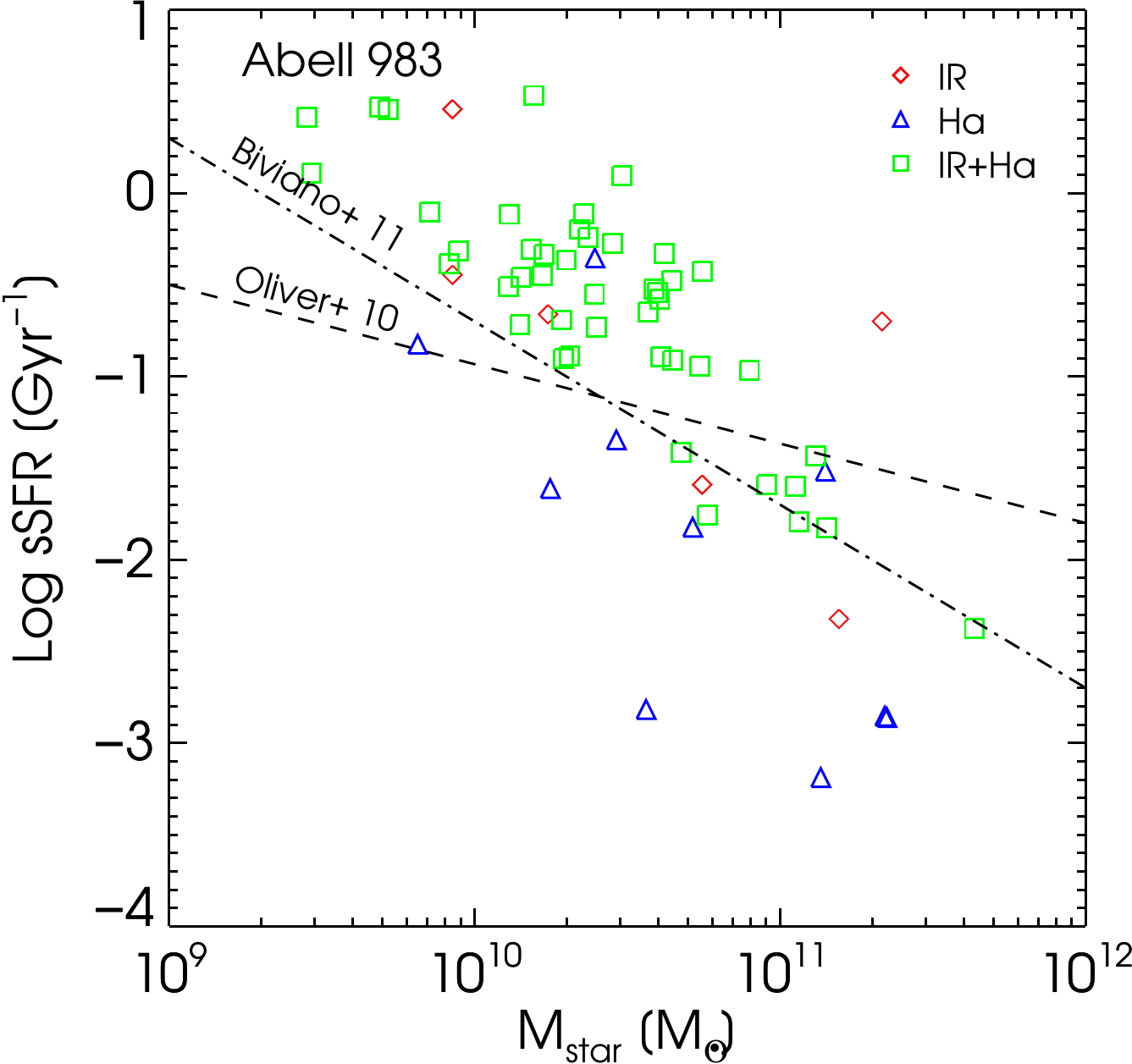}
  \includegraphics[width=0.75\linewidth, keepaspectratio]{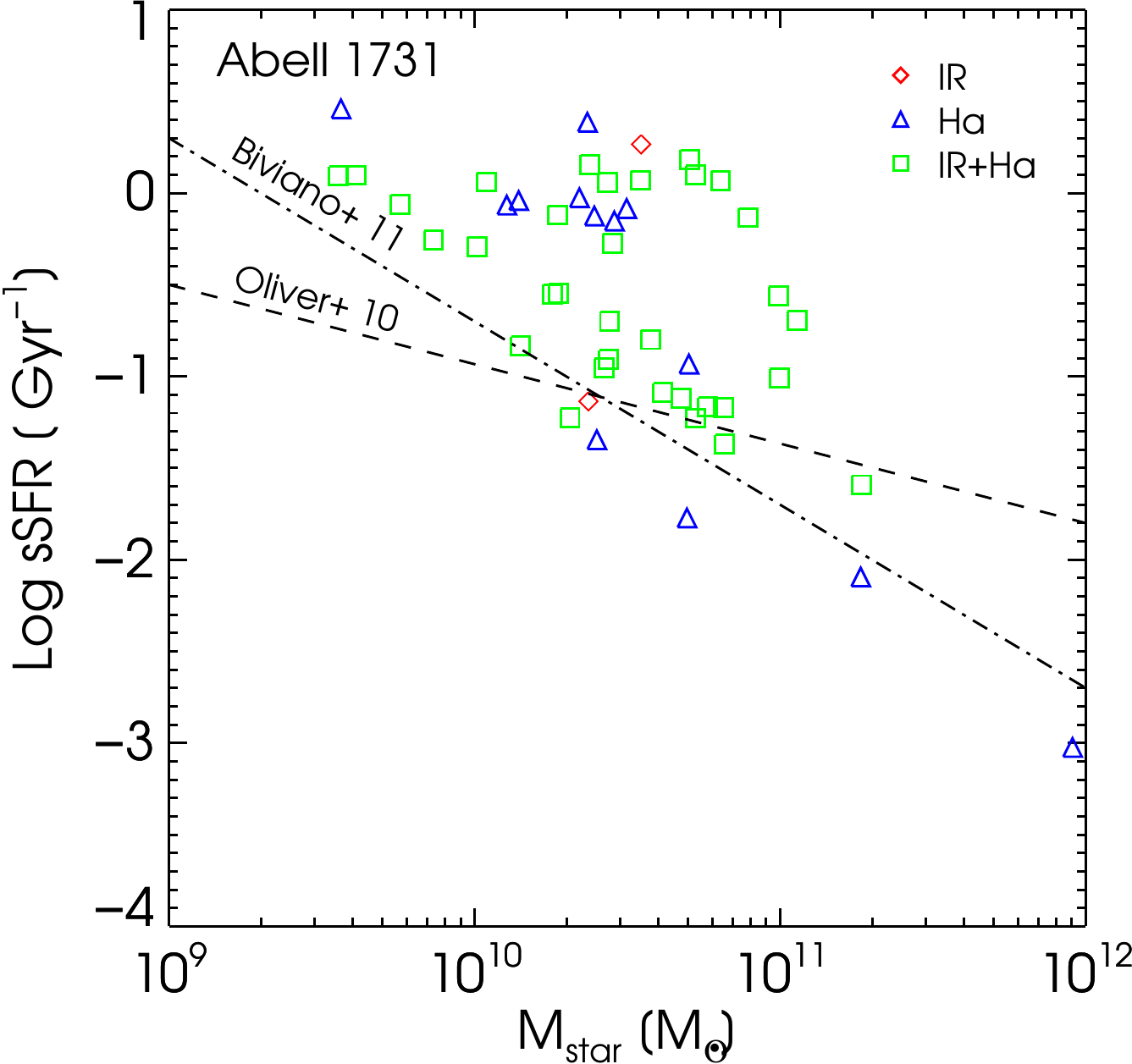}
\caption{The sSFR of the cluster members.  The dot-dashed line
  represent the expected sSFR vs $\rm M_*$ relation for a $24 \mu$m
  source of 0.2~mJy flux density (i.e. at the completeness limit of
  our Spitzer observations), obtained in \citet{biviano11} using the
  relations of \citet{lee10} and \citet{kennicutt98}. The dashed line
  corresponds to the relation of \citet{oliver10} for galaxies from the Spitzer
  Wide-area InfraRed Extragalactic Legacy Survey in the redshift range
  $0.2< z < 0.3$. The galaxies in A983 and 1731 are plotted in
  the top and bottom panel, respectively.}\label{ssfr_comb}
 \end{figure}

\begin{figure}
  \centering
  \includegraphics[width=0.75\linewidth, keepaspectratio]{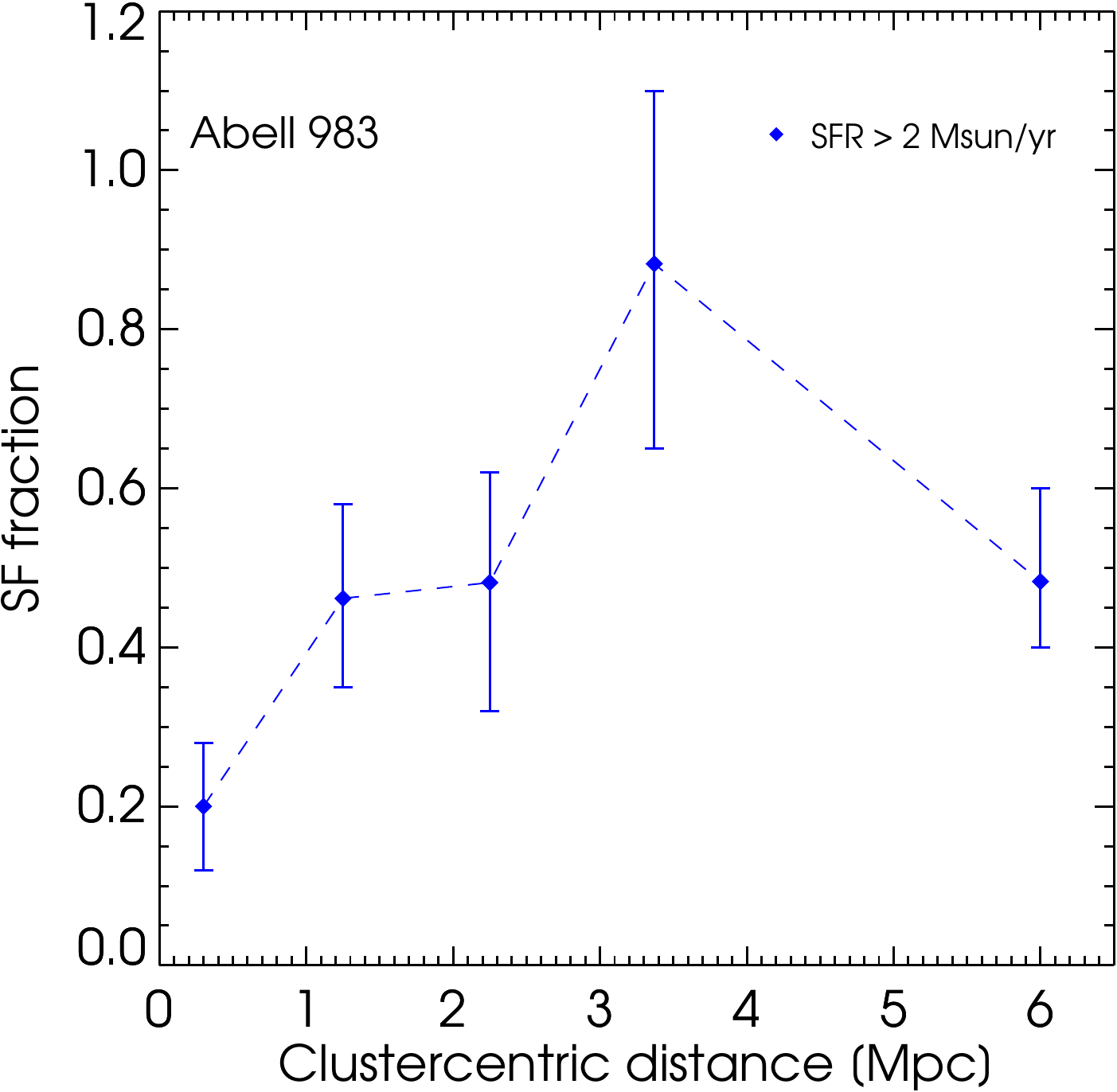}
  \includegraphics[width=0.75\linewidth, keepaspectratio]{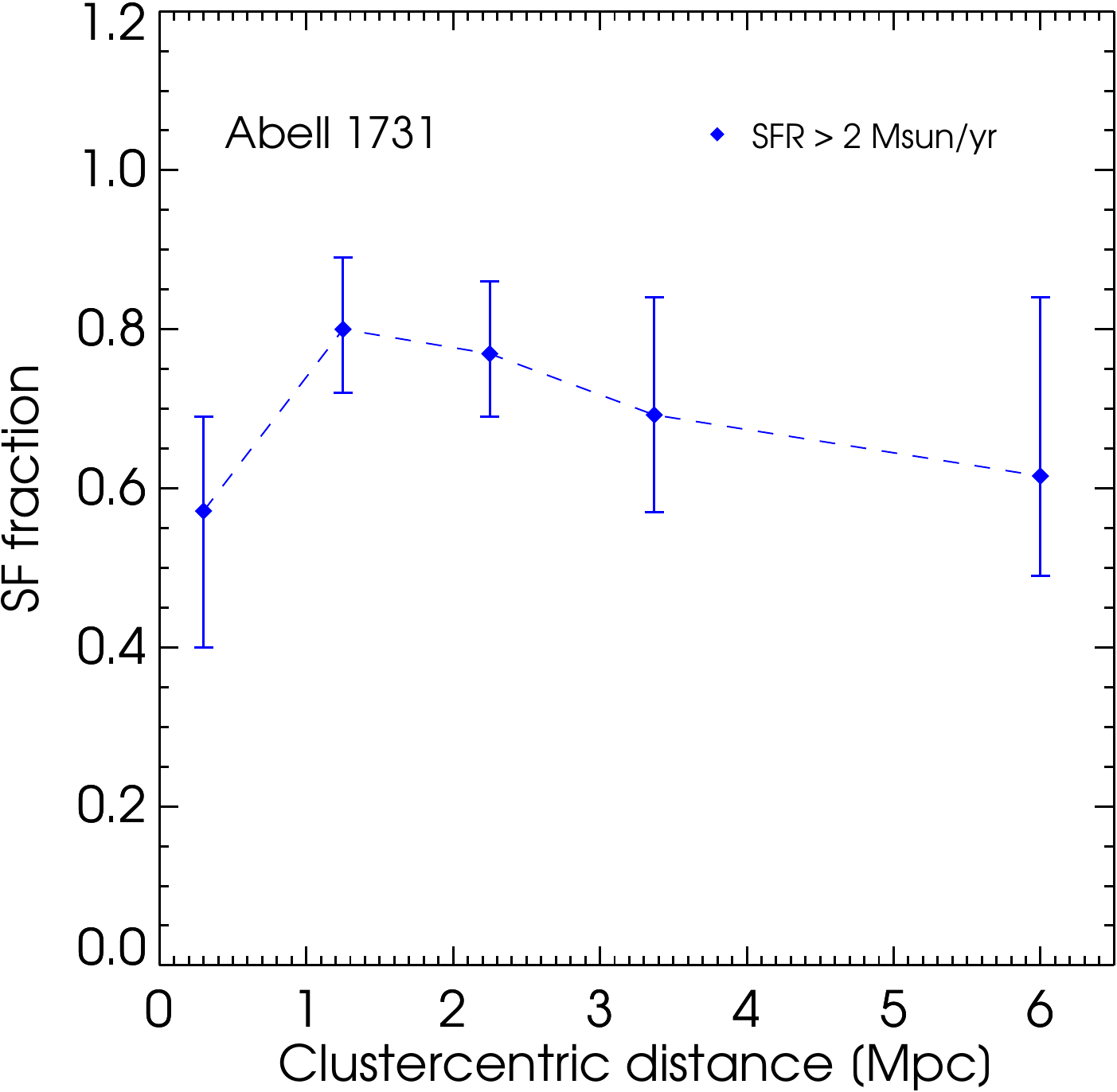}
\caption{The radially binned 1D fraction of star forming galaxies ($\rm SFR> 2\, M_{\odot} \, yr^{-1}$) with respect to the total number of galaxy members, plotted against the clustercentric distance. The errorbars are computed using the bootstrap method. The top and bottom panels refer to A983 and A1731, respectively.}\label{sf-1d}
 \end{figure}

Figure~\ref{ssfr_comb} presents the correlation between sSFR, defined as SFR/$\rm M_{\star}$, and $\rm M_{\star}$. The sSFR quantifies the instantaneous growth rate per stellar mass of a galaxy. We compare our results with the relation
found by \citet{oliver10}. They found a much flatter relation using a
sample of galaxies ($0.2<z<0.3$) from the Spitzer Wide-area InfraRed Extragalactic
Legacy Survey. This difference can be explained considering the
different selection criteria of our work (24 $\mu$m galaxies) and of
\citet{oliver10}, that used a $\rm M_{*}$ selected sample. Therefore, our sample is incomplete at the low mass end. The
dot-dashed line in Figure~\ref{ssfr_comb} represents the average expected
relation between sSFR and $\rm M_*$ for a $24 \mu$m source of 0.2~mJy
flux density, corresponding to the limit below which our sample
becomes incomplete (see \citealt{biviano11}). This limit was computed using the relation of
\citet{lee10} between the $f_{24}$ and $L_{IR}$ at the average
redshift of A983 and 1731, and the \citet{kennicutt98}
relation. For A983, the slope of the relation is close to the one found by \citet{biviano11}. A1731 departs from this trend, presenting a higher
fraction of galaxies with high sSFR. In order to quantify this difference, we selected the galaxies in the mass range $\rm 10^{10}<M_{*} [M_{\odot}] <10^{11}$. We selected this interval as the low mass end of our members sample can be incomplete, and the high mass end is affected by intrinsic low number of objects. We performed a t-test to verify the significance of the higher mean value of the sSFR of A1731 ($\rm Log(sSFR)=0.1\, Gyr^{-1}$) with respect to A983 ($\rm Log(sSFR)=-0.1\, Gyr^{-1}$), finding a low value (0.5) of significance. Therefore, the two distribution are not independent from one another.  Interestingly, we found that $\sim 50\%$ of the galaxies in this mass range are located at intermediate clustercentric distances $\rm\sim 2-3$ Mpc ($\rm \sim1-1.5 \,r_{200}$), for both clusters.
This is compatible with the findings of interacting star forming galaxies at the cluster outskirts by \citet{haines15}. The newly accreted objects can also present on average higher sSFR values with respect to the cluster galaxies as they are not yet fully affected by the environment and hence have conserved the characteristics of the field galaxies \citep{cohen15}. Figure~\ref{sf-1d} shows the 1-D trend of the star forming galaxies, selected to have $\rm SFR\gtrsim 2\, M_{\odot} \, yr^{-1}$, with respect to the total number of members for each cluster. A983 presents a lower fraction of star forming galaxies in the central region, that increases and reaches the maximum value of approximately 90\% at $\rm3-4$ Mpc ($\rm\sim 1.5-2 \,r_{200}$). Similarly, A1731 presents a central dip in the fraction of star forming objects, but the overall distribution is more flat with respect to A983 and has a higher mean value ($\rm\sim60\%$ for A1731 and approximately $\rm\sim45\%$ for A983). We stress that the outskirts of the clusters are suffering from the lower coverage and lower sensitivity of SDSS data, and therefore might be severely incomplete.

\begin{figure}
 \centering
  \includegraphics[width=0.85\linewidth, keepaspectratio]{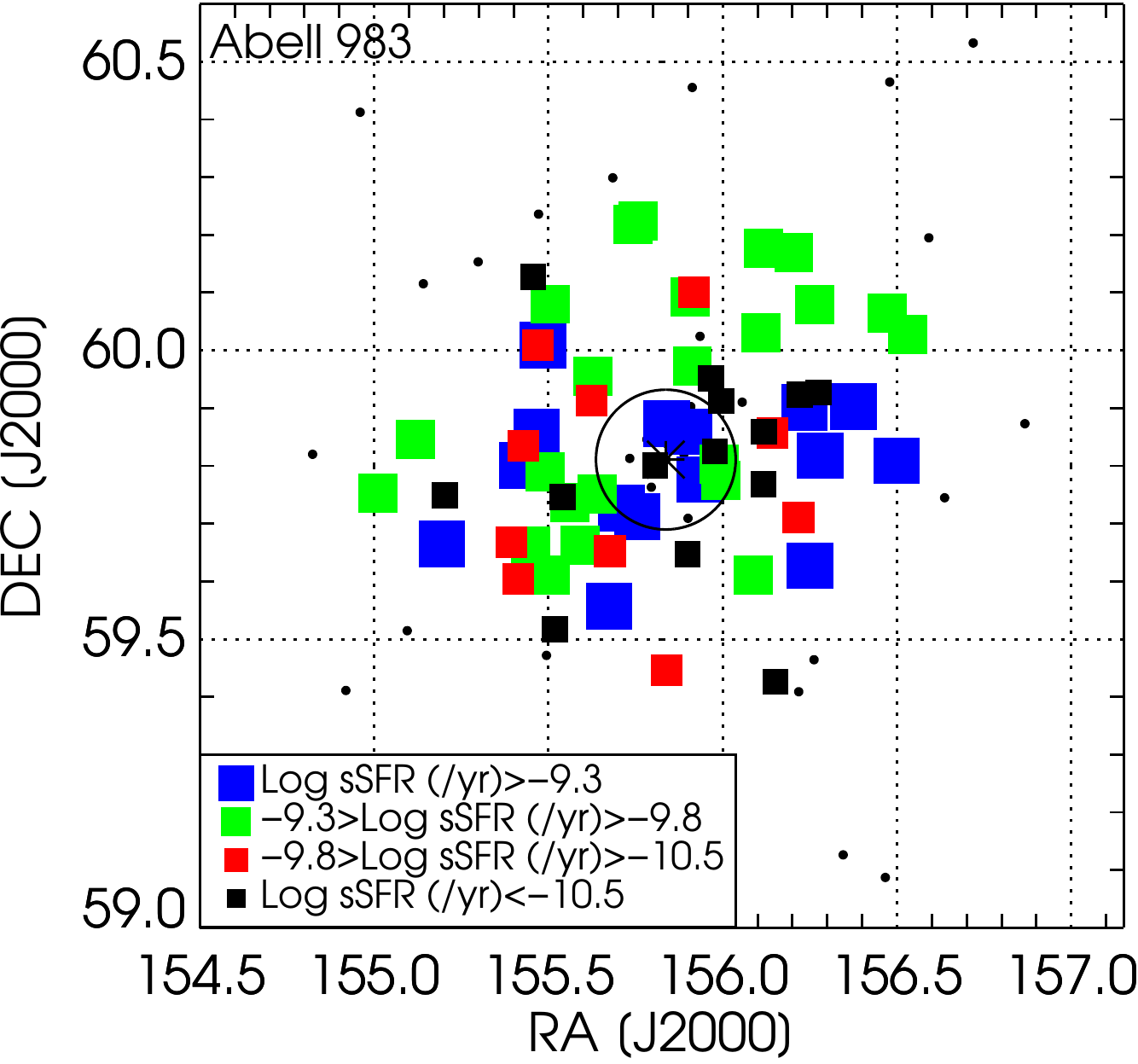} 
  \includegraphics[width=0.9\linewidth, keepaspectratio]{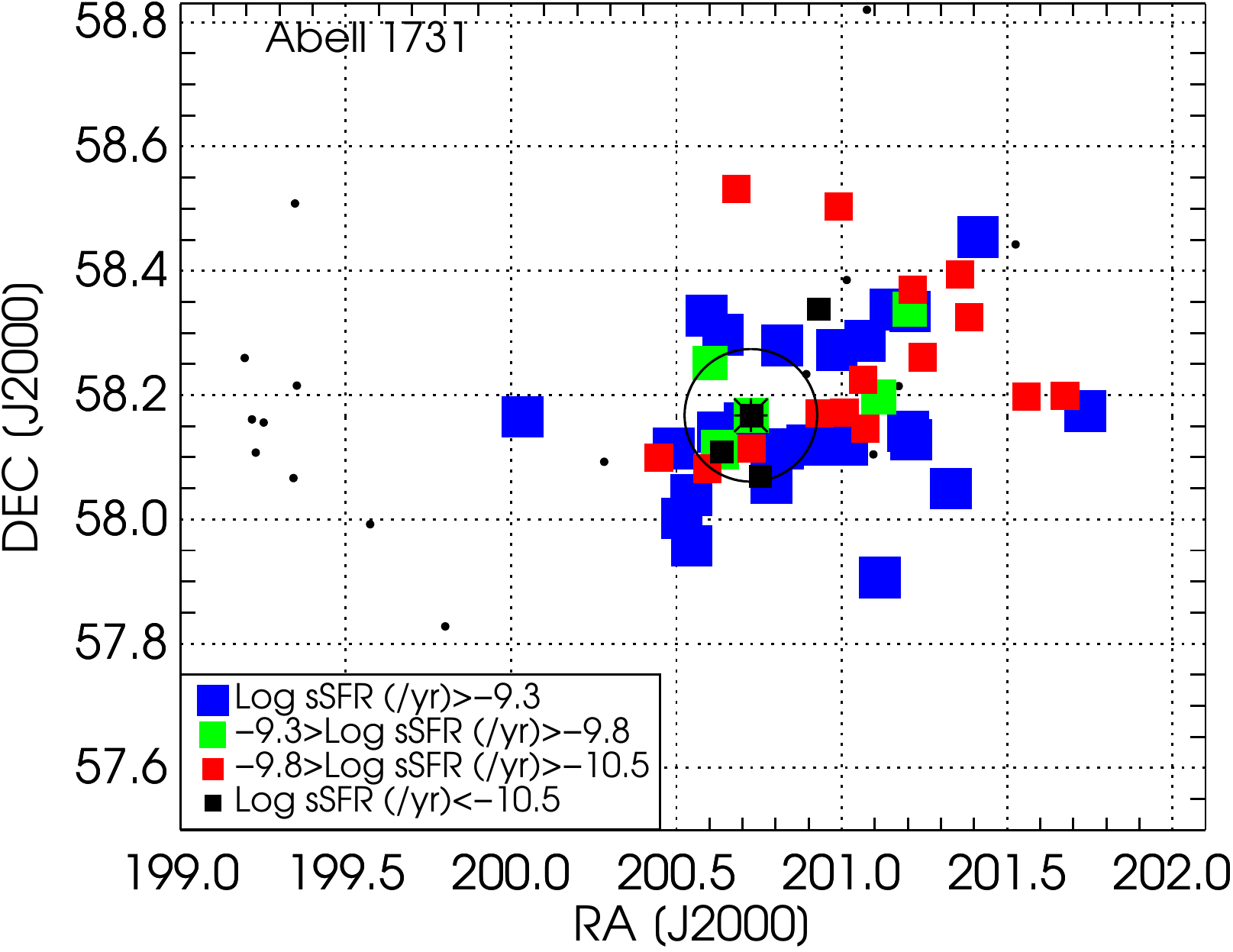} 
  \caption{The 2D distribution of the sSFR of the members in the central region of 
    A983 (top) and A1731 (bottom). The black dots mark the members for which no star formation was measured. The BCG
  is marked as an asterisk and the open cirle marks 1.3 Mpc
  distance from the cluster center.}
  \label{ssfr_comb2d}
 \end{figure}

Figure~\ref{ssfr_comb2d} shows the two dimensional distribution
of the sSFR. A clear asymmetry is present in the spatial distribution of the spectroscopic members of A1731, extending towards the north-west direction. Within this extended structure, $\rm\sim 50\%$ the star forming galaxies present a value of the sSFR in the range $\rm -10.5<sSFR\,[yr^{-1}]<-9.3$, with respect to the remaining fraction that presents $\rm sSFR [yr^{-1}]\,>-9.3$.
\citet{fadda06} and \citet{edwards10} pointed out that accreted galaxies
undergo an accelerated evolution when they meet the denser cluster
environment, experiencing a short starburst phase followed by the
quenching of their star formation activity. \citet{cohen14} and
\citet{cohen15} find a higher fraction of star forming galaxies in
dynamically active cluster than in more relaxed ones, by measuring
star formation with optical spectroscopic data from SDSS and infrared
data from WISE. 
The lower sSFR in A1731 in half the galaxies at these intermediate clustercentric distances ($\rm\sim1.5-2 \,r_{200}$) could be due to the on-going harassment process, that leads to the consumption of the gas reservoir in the infalling galaxies and the reduction of the sSFR. This extended structure presents the spatial characteristic and the pre-processing evidence \citep{haines15} of a filament-like structure. Additional spectroscopic observation and X-ray imaging would complement the currently available dataset, highlighting the presence of infalling groups of galaxies or a filament.
  
 \begin{figure}
 \centering
  \includegraphics[width=0.7\linewidth, keepaspectratio]{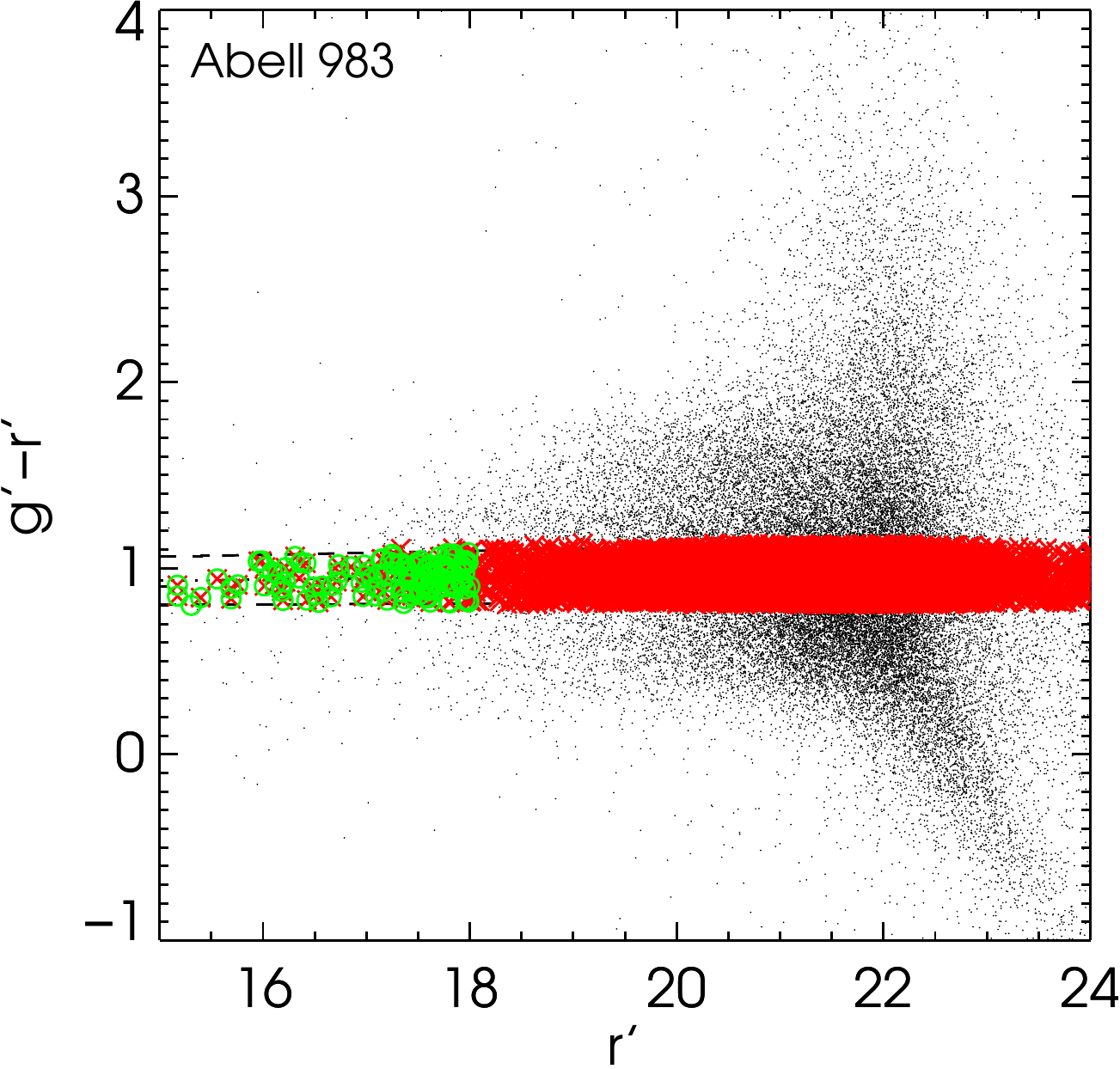}
  \includegraphics[width=0.7\linewidth, keepaspectratio]{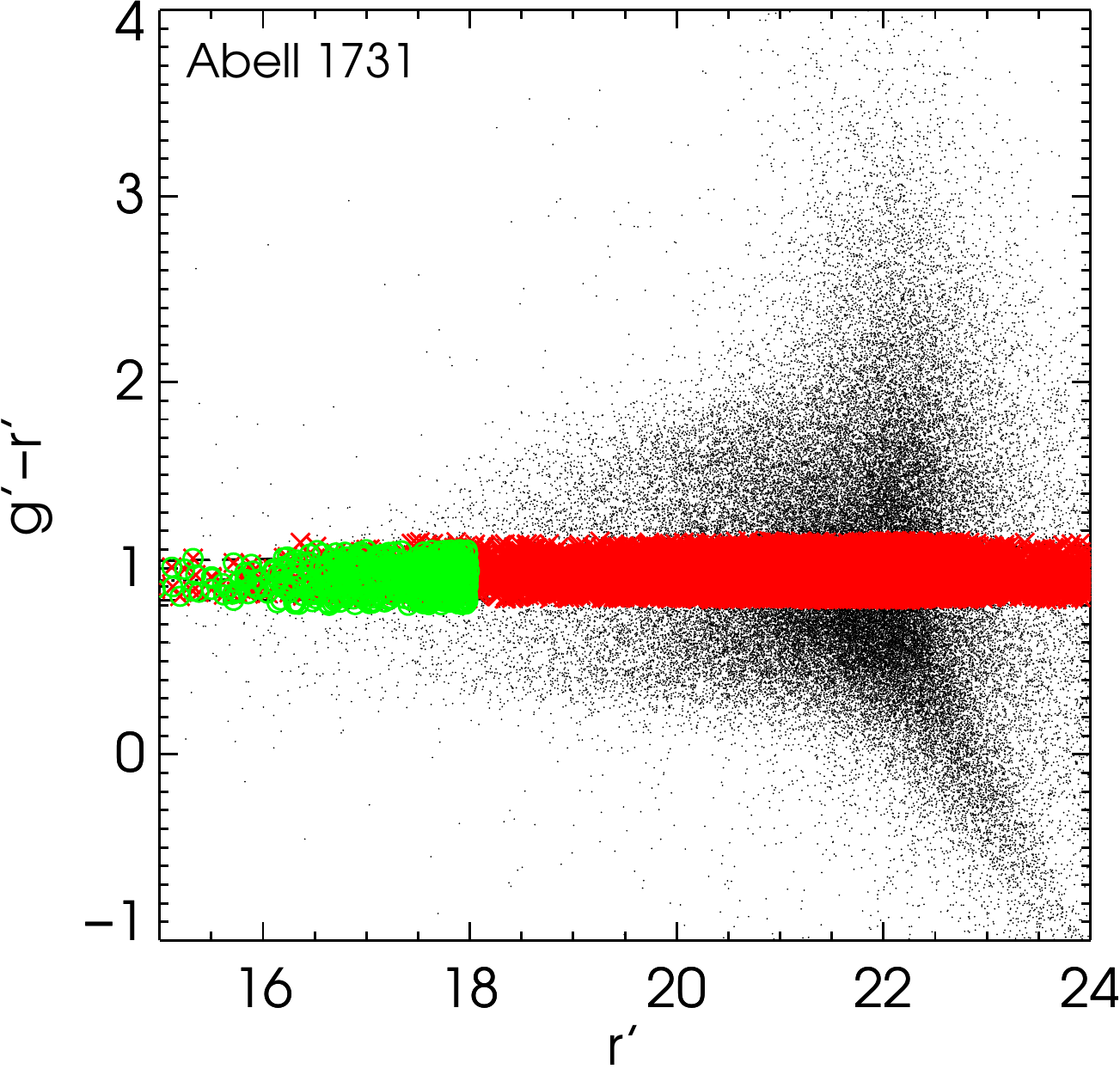} 
\caption{The color magnitude diagram of the cluster A983 (top
  panel) and 1731 (bottom panel). The galaxies belonging to the red
  sequence are marked with red crosses. The subsample of the galaxies
  used for the red-sequence selection is visible as green circles.}\label{cmd}
 \end{figure}  
 
    \begin{figure}
 \centering
  \includegraphics[width=\linewidth, keepaspectratio]{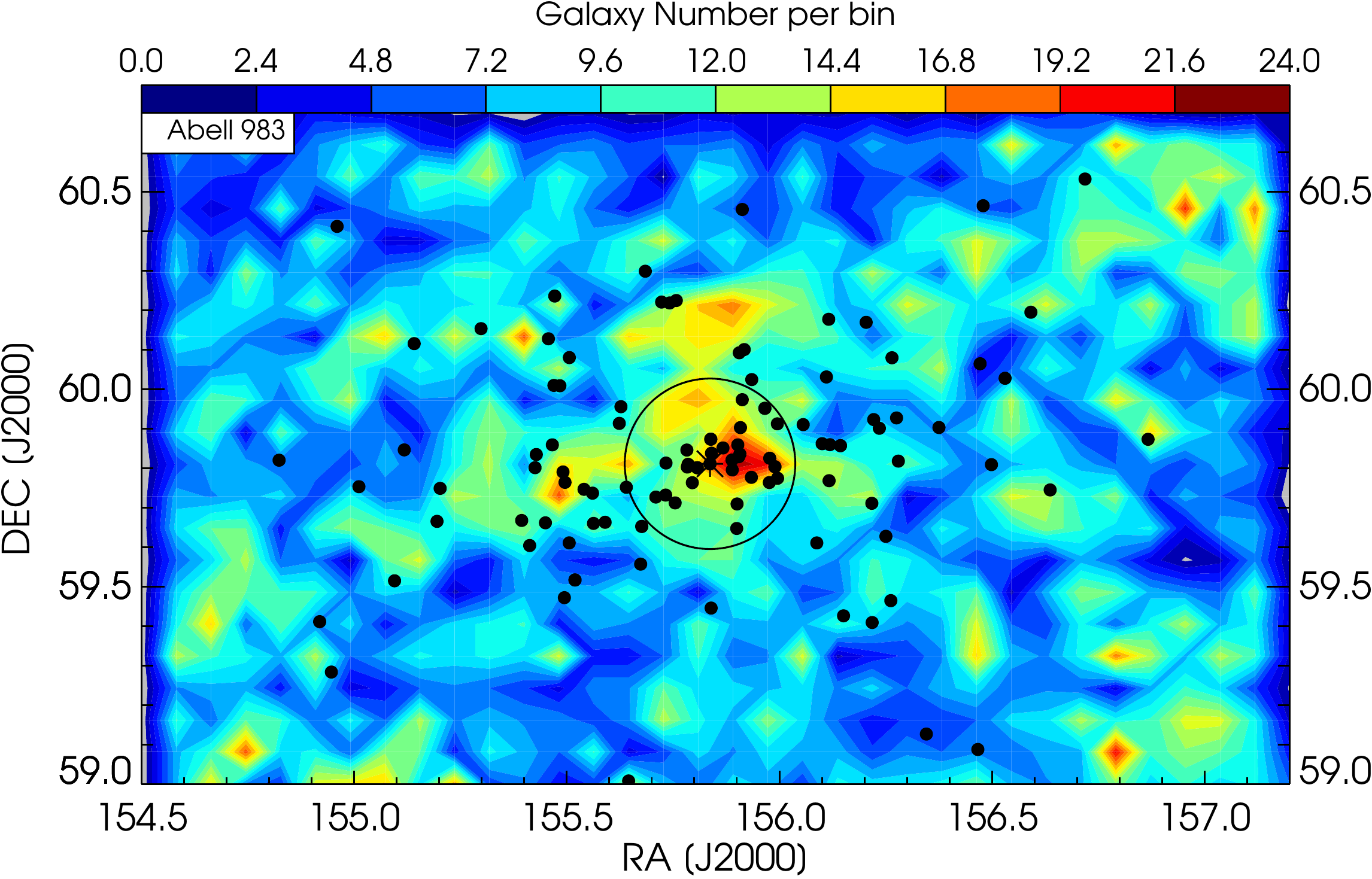}
  
  \includegraphics[width=\linewidth, keepaspectratio]{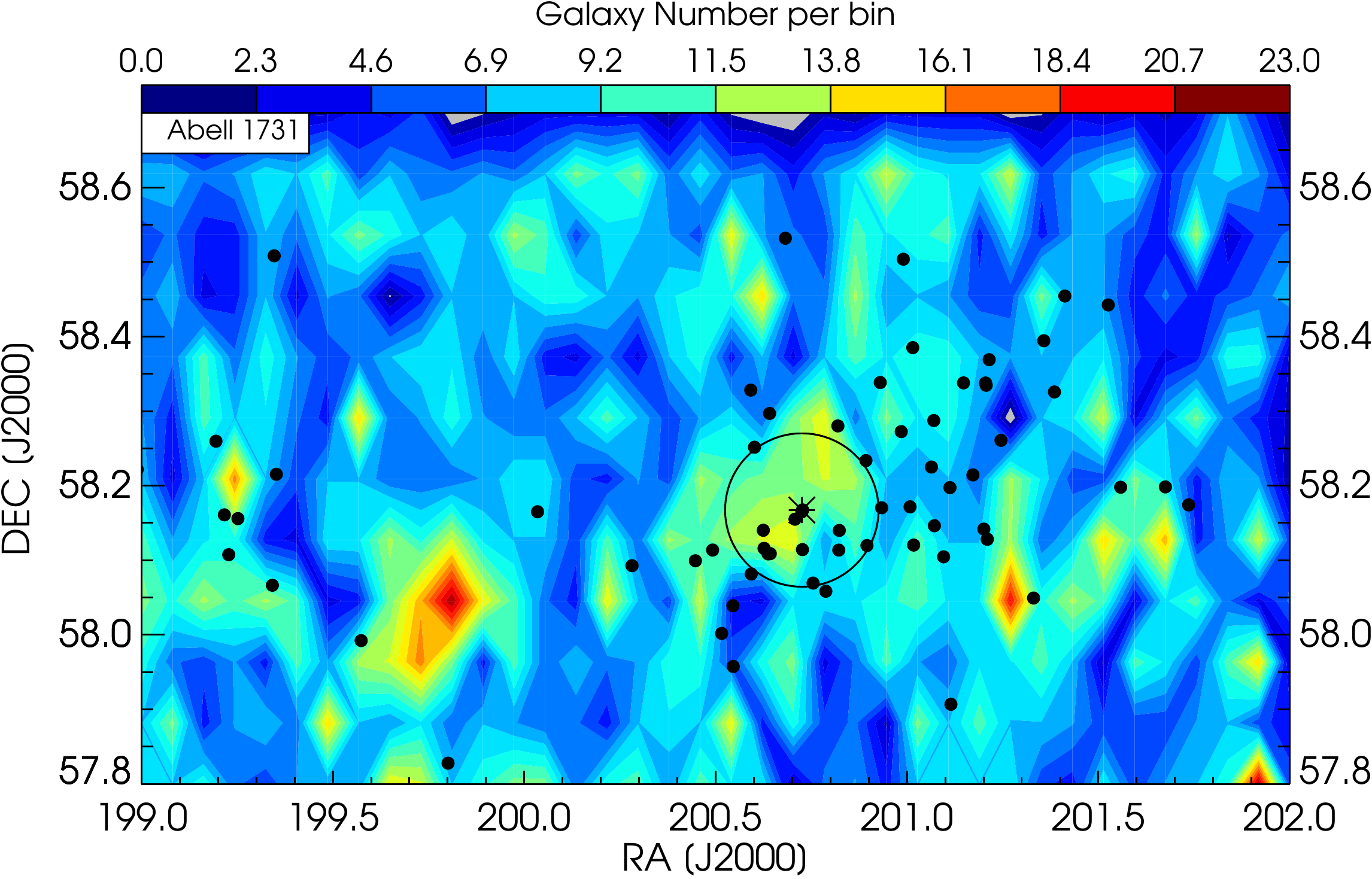} 
\caption{The 2D density distribution of the passive galaxies  vs the dynamical center of the clusters (plotted as an asterisk). The spectroscopically confirmed members are
  overplotted as black circles. The panels for A983 and 1731 are the top and bottom one, respectively. the open cirle marks 1.3 Mpc
  distance from the cluster center.}\label{passive_dynamic}
 \end{figure}

As an additional comparison, we built the color magnitude diagram of
the g'-r' versus r' magnitude from SDSS DR10 of the galaxies in the
same region of the sky with respect to the clusters (Figure~\ref{cmd}). The passive
population of the member galaxies falls in what is called a
red-sequence, clearly separated from the more star forming and
therefore bluer cloud of field galaxies. We visually selected the
galaxies in the red-sequence down to a $\rm r'=18$. We performed a
linear fit of this subsample, and extended the relation to the whole
sample, applying a 1$\rm\sigma$ error cut above and below the fitted
line. The selected passive population of galaxies is visible in
Figure~\ref{cmd}. We compared the density distribution  of the red-sequence galaxies to the distribution of the galaxies selected via the shifting gap method (see Figure~\ref{passive_dynamic}). Our aim is to verify the match between the evolutionary and dynamical centre of the clusters. A983
presents a direct match between the evolutionary and the dynamical centre (offset $\sim 1'$).  A1731 shows a similar offset ($\sim 2'$),
although presenting less passive galaxies in the central 1 Mpc region ($270$ and $120$ galaxies in A983 and 1731, respectively). The extended structure showed in Figure~\ref{ssfr_comb2d} does not correspond to any overdense region of the passive galaxies. The scenario of an actively accreting cluster
appears therefore the most likely.  The dense clump in Figure~\ref{passive_dynamic} located 
 at $\rm R.A.=201.3$ $\rm Dec.=58.1$ presents no clear identification. The
NED archive suggests the presence of a galaxy group in proximity of
those coordinates \citep{mcconnachie09}. The
overdensity at $\rm R.A.=199.7$ $\rm Dec.=58.0$ corresponds to the cluster Abell~1713,
located at z$\sim$0.14.

\subsection{The effect of the dynamical state of the cluster on the AGN population}

\begin{figure}
 \centering
  \includegraphics[width=0.75\linewidth, keepaspectratio]{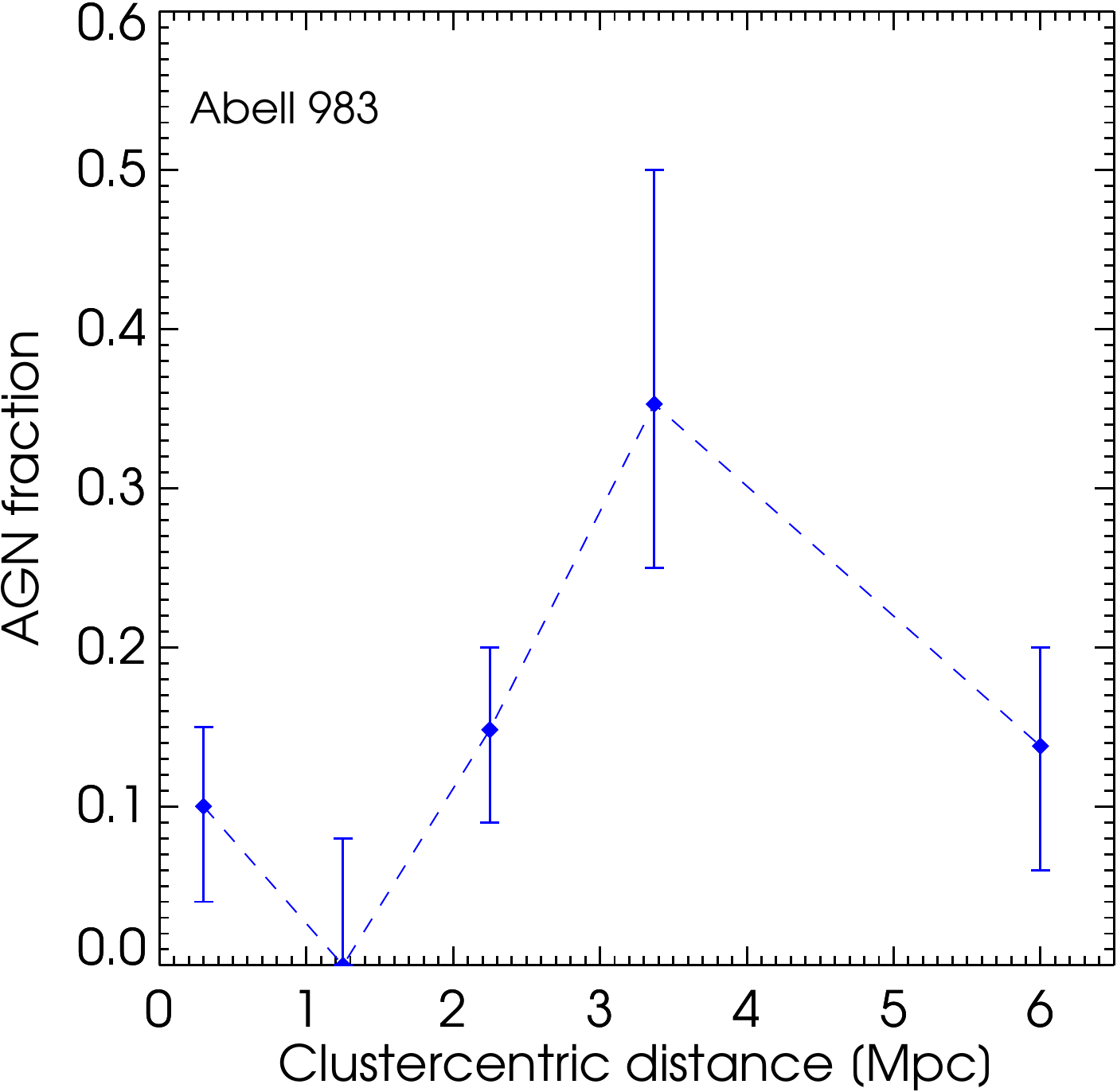}
  \includegraphics[width=0.75\linewidth, keepaspectratio]{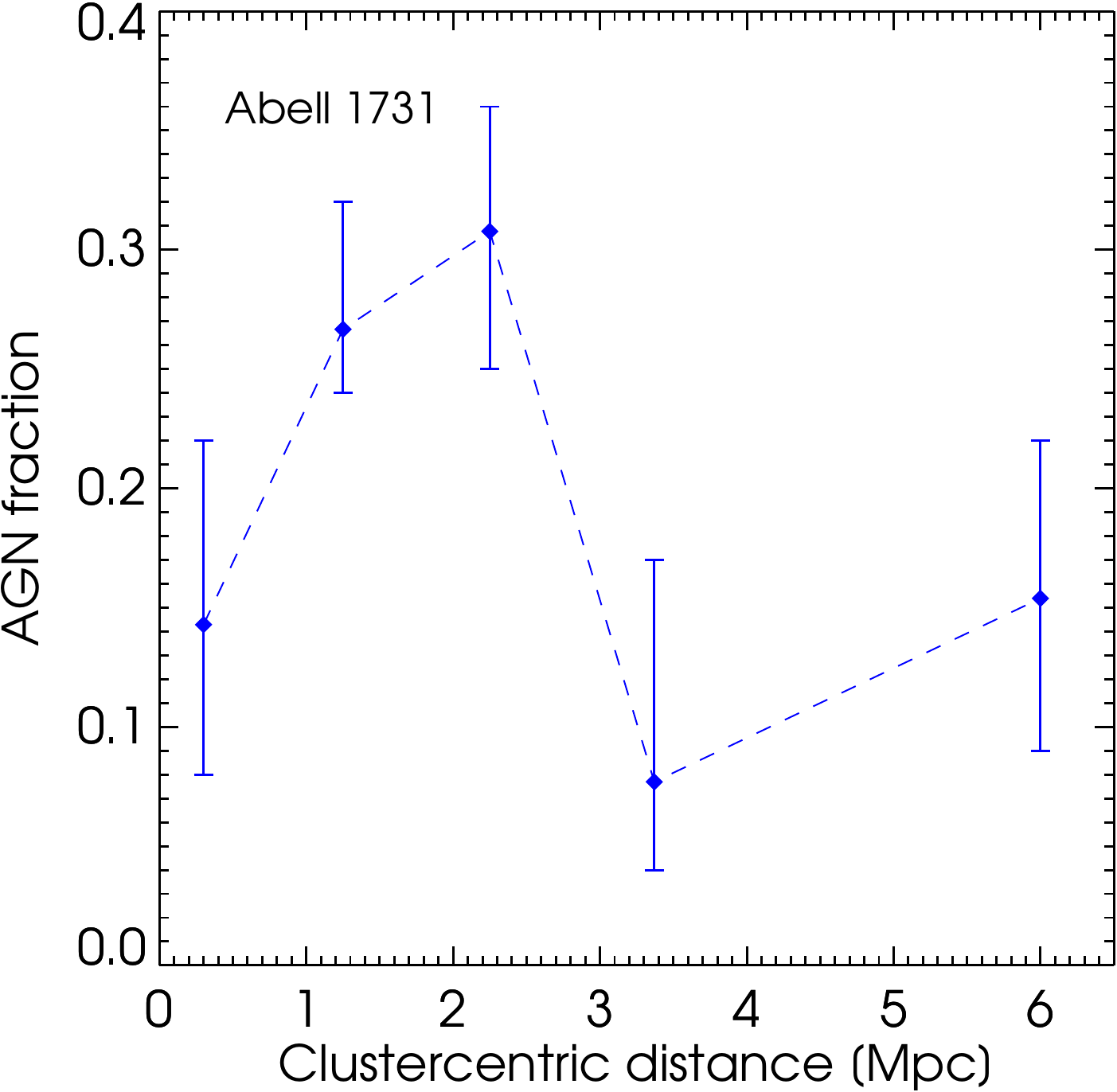} 
\caption{The radially binned 1D distribution of the AGN fraction plotted against the clustercentric distance, for A983 (left panel) and A1731 (right panel). The errorbars are computed using the bootstrap method.}\label{ir_fraction2d}
 \end{figure}

Gravitational and hydrodynamical interactions are known to cause the disturbance of the gas content in galaxies. The compression and collapse of the gas in the central region of a galaxy can trigger AGN activity (\citealt{springel05}, \citealt{hong15})
On the other hand, ram pressure stripping,
although more efficient in depleting the outskirts of galaxies as they
enter the cluster environment, could influence the inner reservoir of gas
that powers the AGN. In
Figure~\ref{ir_fraction2d}, both clusters show a dearth of AGN
 in the cluster centre (1 Mpc, $\rm \sim0.5 \,r_{200}$,
from the BCG), followed by an increase at the outskirts ($\rm 2-4$ Mpc, $\rm \sim 1-2 \,r_{200}$,
from the BCG). At higher clustercentric distances (larger than 5 Mpc,  $\rm \sim 2.5 \,r_{200}$, from the BCG), the clusters present a different behavior, with A983 showing a steep decrease of the AGN fraction, whereas A1731 shows a slight increase (from 10\% to 16\%). Both clusters present no significant difference when
comparing the cluster central and distant, field-like, regions, but a
steep increase in the number of AGN is found on average at about 2 Mpc ($\rm \sim 1 \,r_{200}$)
from the BCG. This result is in agreement
with the results of \citet{haines13} and \citet{pimbblet13}. At intermediate radii, the frequent gravitational interactions of gas rich galaxies might favour the growth of instabilities in the gas distribution. These  are accreted on the central black hole, causing an increase of its activity. Our
study confirms that the majority of the active AGN are found at
intermediate ($\rm \sim 1-2 \,r_{200}$) clustercentric distance, where the infalling galaxies
have still to be fully processed in the cluster environment. This holds also
when considering the non symmetrical distribution of the members in A1731, indicating a relation between the AGN fueling efficiency and the global density of the 
ICM. 
The dual effect of the cluster environment on the AGN life cycle appears confirmed. On the local scale, the high density of galaxies promotes the instability and accretion of gas on the central black hole. On the large scale, the interaction with the ICM inhibits the subsequent replenishment of the gas reservoir.

\section{Summary and Conclusions}\label{summary}

We surveyed  the activity of the member galaxies in A983 and 1731,
focusing on their star formation and black hole accretion. Both clusters are located at
z$\sim$0.2 and their members were selected using
spectroscopic redshifts. The total IR luminosity and $\rm M_*$ of the
members were computed using SED fitting. The total IR luminosity
was used to compute the obscured SFR of the galaxies. Furthermore, the
spectroscopic follow-up allowed us to obtain redshifts and independent
unobscured SFR measurements, using the $\rm H_{\alpha}$ line flux. AGN
were identified using an IR color cut, narrow line flux ratio and
broad-line detection, and removed from the SF analysis.

The distribution of passive population of galaxies presents a density peak in the central region for both clusters, albeit the number of passive objects is twice in A983 than in A1731. A clear difference emerged when
comparing the 2D distribution of the star forming galaxies:
A983 presents symmetric distribution of star forming members, whereas those in
A1731 were extended
along what appeared to be a filament-like structure. 
Higher values for the sSFR were
found in both cluster outskirts, at $\rm 2-3$ Mpc ($\rm \sim 1-1.5\,r_{200}$) in clustercentric distance .  
A983 showed the typical characteristics of an evolved cluster. In contrast, A1731 appeared to be undergoing a phase of active accretion of field galaxies. The $50\%$ of members hosted in the extended structure in A1731 presented lower values of the sSFR  with respect to the other members at the same clustercentric distance. This suggested the possible effect of galaxy-galaxy interactions in reducing the sSFR, via harassment and subsequent suffocation, as expected due to the pre-processing of galaxies in transitional environment, such as filaments or galaxy groups.

The activity of AGN depends on the availability of cold gas in the central regions
of the galaxy, and on the dynamical processes that influence its
accretion on the black hole.
The diagnostics we used allowed us to observe a radial trend of the
presence of AGN, that were on average found at intermediate
clustercentric distances ($\sim$3 Mpc, $\rm \sim 1.5 \,r_{200}$), where galaxy-galaxy interactions are more frequent (\citealt{balogh04}, \citealt{gallazzi09}). These results are in
agreement with \citet{haines13} and \citet{pimbblet13}.  
The cluster environment appears to have an influence on a dual scale: on the small scale, gas instabilities are promoted due to the frequent gravitational and hydrodynamical interactions, leading to new SF and black hole accretion episodes; on larger scale, further accretion of cold gas on
the galaxies is blocked, suffocating the star formation and AGN activity. 
Further investigations, using both observations and simulations are necessary.  Additional spectroscopic observations and X-ray imaging are required to unmistakably identify the presence of infalling groups or filaments in A1731.  This would help in disentangling the effects of physical processes responsible for the accelerated ageing
of cluster members.

\section{Acknowledgements}
The authors would like to thank the anonymous referee whose suggestions improved the quality of the manuscript.
The authors would like to thank Tom Jarrett for kindly providing his package for the reduction of the WIRC near-IR data. The authors would like to thank Andrea Biviano for kindly providing their code \citep{mamon10} for the cluster membership selection.

\bibliographystyle{aa}
\bibliography{biblio.bib}

\end{document}